\documentclass[prb,aps,onecolumn,showpacs,floats,a4paper,times,12pt]{revtex4}
\usepackage{epsfig,bm,epsf,graphics}
\usepackage{xcolor}
\newcommand{\La}{\mathop{\rm La}}
\newcommand{\Sr}{\mathop{\rm Sr}}
\newcommand{\Mn}{\mathop{\rm Mn}}
\newcommand{\Oks}{\mathop{\rm O}}
\newcommand{\Ti}{\mathop{\rm Ti}}
\newcommand{\Pb}{\mathop{\rm Pb}}
\newcommand{\Zr}{\mathop{\rm Zr}}
\newcommand{\R}{\mathop{\rm R}}
\newcommand{\Fe}{\mathop{\rm Fe}}
\newcommand{\Ho}{\mathop{\rm Ho}}
\newcommand{\Tb}{\mathop{\rm Tb}}
\newcommand{\Sn}{\mathop{\rm Sn}}
\begin{document}
%\begin{frontmatter}
\title{Magneto-Ferroelectric Interaction in Superlattices: Monte Carlo Study of Phase Transitions}
\author{I. F. Sharafullin $^{a,b}$, M. Kh. Kharrasov $^{b}$, H. T. Diep\footnote{Corresponding author, diep@u-cergy.fr} $^{a}$}
\address{$^{a}$ Laboratoire de Physique Th\'eorique et Mod\'elisation,
Universit\'e de Cergy-Pontoise, CNRS, UMR 8089, 2 Avenue Adolphe
Chauvin, 95302 Cergy-Pontoise, Cedex, France.\\
 $^{b}$ Bashkir State University, 32, Validy str, 450076, Ufa, Russia.}

\begin{abstract}
We study in this paper the phase transition in superlattices formed by alternate magnetic and ferroelectric layers, by the use of Monte Carlo simulation. We study effects of temperature,
external magnetic and electric fields, magnetoelectric coupling at the interface
on the phase transition.
 Magnetic
layers in this work are modeled as thin films of simple cubic lattice with Heisenberg
spins. Electrical polarizations of $\pm{1}$ are assigned
at simple cubic lattice sites in the ferroelectric layers. The transition temperature, the layer magnetizations,
the layer polarizations, the susceptibility, the internal energy, the
interface magnetization and polarization are calculated.
The layer magnetizations and polarizations as functions of temperature are shown for
various coupling interactions and field values.
Mean-field theory is also presented and compared to MC results.
\end{abstract}

\maketitle

\section{Introduction.}\label{sect-intro}

The study of phase transitions, surface effects and critical
phenomena in superlattices or multilayered magnetic nanofilms has
been rapidly developed during the last two decades (see reviews
\cite{diep:hal-01084599,Diep201631,Prudnikov2015,Ramazanov2016,
kamilov1999}). Such high interest in this area was stimulated by the fact that superlattices of nanofilm and multiferroics possess a
number of unique properties which have a broad area of applications in
nanoelectronics, spintronics \cite{Diep201631,pyatakov,
PhysRevLett,Iijima_1992,Oneill2000,PhysRevB.55.11218,ramesh} and
devices using the giant magnetoresistance phenomenon
\cite{diep:hal-01084599,Prudnikov2015,PhysRevB.85.184413}.

With modern technologies it is possible to create superlattices and
multilayer nanofilms as thin as a few atomic layers from the crystal
structures with magnetic and ferroelectric orderings. These
structures are able to manifest magnetoelectric effects which are
known to be the result of interactions between magnetic and ferroelectric subsystems.
 It should be noted that the study of
magnetoelectric effects in these systems draws a great fundamental
interest for their special features, such as size dependence of
magnetic and ferroelectric order parameters and other
characteristics \cite{pyatakov,PhysRevLett,Kharrasov2016,PhysRevB.73.094434}. For example, it has been shown that the change from the
bulk values for films of a few dozens of monolayers ($d\geq 10$ nm) to
the two-dimensional values for films thinner than 4-6 monolayers ($d
\leq(1-2)$ nm) \cite{Prudnikov2015,Kharrasov2016}.

In Ref. \onlinecite{lamekhov2015} it was shown that in heterostructures with
magnetic and ferroelectric materials, the magnetoelectric effect
induced by an external electric field is observed at the interface
layer. This effect is accompanied by the appearance of an
antiferromagnetic phase at the interface as well as with the change
in the critical temperature of the magnetic layer. This has been
observed experimentally in
$\La_{0.87}\Sr_{0.13}\Mn\Oks_{3}/\Pb\Zr_{0.52}\Ti_{0.48}\Oks_{3}$
\cite{PhysRevB.87.094416} at certain concentrations of
manganite. In the work Ref. \onlinecite{Ort2014} with Monte Carlo (MC) simulation
for a two-layer film with the structure
$\La\Sr\Mn\Oks/\Pb\Zr\Ti\Oks$, phase transitions have been
investigated and the correctly describing model has been
proposed. A multi-sublattice model has been introduced to explain
magnetic properties of compounds $\R_{2}\Fe_{17},
\Ho_{2}\Fe_{11}\Ti$ and $\Tb\Mn_{6}\Sn_{6}$
\cite{0953-8984-14-27-310,PhysRevB.66.014437,PhysRevB.55.12408}.

On theoretical points of view, one of the most studied systems for the
layered magnetic structure was concentrated on the magnetic
properties of magnetic bilayer \cite{Diepbook2}. Wei Wang et al. \cite{WANG2017104}
have studied a ferrimagnetic
mixed spin (1/2, 1) Ising double layer superlattice: they have shown the effects of the exchange coupling and the layer thickness
on the compensation behavior and magnetic properties of the system, by
MC simulation. Some interesting phenomena have been found,
such as various types of magnetization curves, originating from the
competition between the exchange coupling and temperature. In Ref. \onlinecite{Fer2016} the phase diagram and magnetic
properties of the mixed spin (1, 3/2) Ising ferroelectric
superlattices with alternate layers have been investigated by means
of MC simulation. It should be noted that they also
investigated superlattice of only two ferroelectric layers with
antiferroelectric interfacial interaction between layers, within the
transverse Ising model. They found a number of interesting
phenomena, such as the existence of the compensation temperature or
transverse field to compensate the specific ranges of exchange
interactions.

Note that MC methods based on
the Metropolis algorithm, as well as other algorithms have proven to
be successful in describing physical properties of magnetic
systems of different spatial dimensions. They revealed particular features
of the phase transitions in these systems
\cite{landau2014guide}. In Refs. \onlinecite{Prudnikov2015}, \onlinecite{phu2009crossover}, \onlinecite{phu2009critical} and \onlinecite{PhysRevB.91.014436} numerical studies of size effects in critical
properties of the Heisenberg multilayer films with MC
methods were conducted. For films of varying thickness an
anisotropy induced, for example, by
the crystalline field of the substrate, was taken into account. The precise calculation of
critical indices was carried, the values of which have clearly
demonstrated the dimensional transition from two-dimensional to
three-dimensional properties of the films with increasing number of
layers.

In this paper the methodology for Heisenberg
multilayer films simulations \cite{diep:hal-01084599,landau2014guide} is used for the MC simulation and the
calculation of magnetic properties of multiferroic superlattices.
Although previous valuable theoretical, numerical and experimental
studies have been done, more research is still needed to further
understand the magnetic, ferroelectric and thermodynamic properties
of the superlattice. Our investigation is motivated by the fact
that superlattices with magnetic and ferroelectric materials present great opportunities of applications
in spintronics.

\begin{figure}[h]
\vspace{10pt}
\begin{center}
\includegraphics[scale=0.33]{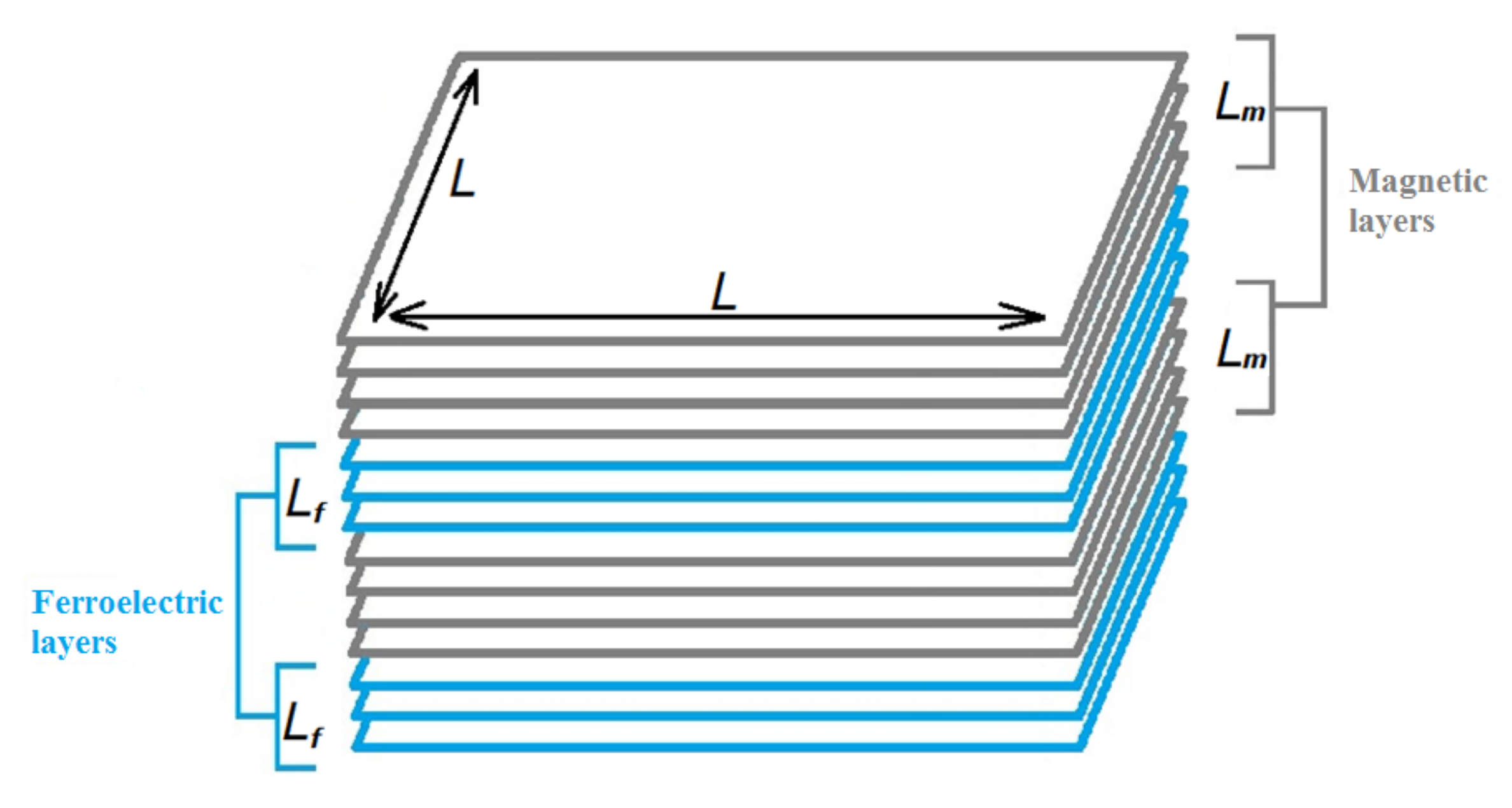}
\end{center}
\vspace{10pt}
\caption{Schematic representation of the superlattice: alternate
ferroelectric and magnetic films.}\label{ref-fig1a}
\vspace{20pt}
\end{figure}

In the present paper, we will thus study the effects
of the magnetoelectric coupling and the external magnetic and
electric fields on the magnetic properties of the multiferroic
superlattice shown in Fig. \ref{ref-fig1a}.

%Fig1

The paper is organized as follows. The model of the superlattice is presented in section~\ref{sect-model}, where we summarize the principal
steps used in the calculation of the ground-state configurations of
the system.
Section~\ref{sect-results} shows the MC results of energy,
layer magnetizations, susceptibilities and layer polarizations.
Section \ref{other} shows results of another choice for interface coupling. The mean-field (MF) theory is shown in section \ref{MF}. Concluding remarks are given in section~\ref{sect-conc}.

\section{Model and Ground State}\label{sect-model}

\subsection{Model}
We  consider a multilayer multiferroic
films  composed of $L_z^m$ ferromagnetic layers and
$L_z^f$ ferroelectric layers alternately sandwiched in the $z$ direction (see Fig.~\ref{ref-fig1a}). Each $xy$ plane has the dimension $L\times L$.
 The lattice sites of the magnetic layers of this superlattice are occupied by interacting Heisenberg spins $\vec{S}$, while the lattice sites of the ferroelectric layers are occupied by interacting polarizations $\vec{P}=\pm 1$ along the $z$ axis. Our system thus consists of a $L\times L\times L_z $ sites where
 $ L_z=L_{z}^m+L_{z}^f$.  We assume periodic boundary conditions in all directions to reduce surface effects.
We assume interactions between
ferroelectric and magnetic systems at their interfaces. The Hamiltonian
of the system is defined as follows:

\begin{equation}\label{eq-ham-sys-1}
{\cal  H}=H_{m}+H_{f}+H_{mf},
\end{equation}

The first term is the Hamiltonian of the magnetic subsystem,
the second - of the ferroelectric subsystem, the third term is the
Hamiltonian of their interaction. We assume

\begin{equation}\label{eq-ham-sysm-1}
H_{m}=-\sum_{i,j}{J^{m}_{ij}
\vec{S}_{i}\cdot \vec{S}_{j}}-\sum_{i}(\vec{H}\cdot\vec{S}_{i})
\end{equation}
here $J^{m}_{ij} >0$ characterizes the ferromagnetic interaction
between one spin and its nearest neighbors (NN). We consider it to be the
same for NN within a layer and NN in adjacent layers. $\vec{S}_{i}$ is the classical Heisenberg spin occupying the i-th site. $\vec{H}$
is an applied magnetic field along the $+z$ direction.
For the ferroelectric subsystem we
write
\begin{equation}\label{eq-ham-sysf-1}
H_{f}=-\sum_{i,j}{J^{f}_{ij}P_{i}\ P_{j}}-\sum_{i}{P}_{i}\ E^z
\end{equation}
where $P_{i}$ is the polarization along the $z$ axis at the i-th site assumed to have only two values $\pm 1$ (Ising-like model),
$J^{f}_{ij} >0$ denotes the NN ferroelectric interaction, similar for all NN.  $E^z>0$ is the external
electric field applied along the $+z$ axis perpendicular to
the plane of the layers.

The magnetic interface layer creates at a site $k$ of the ferroelectric interface an
effective field $H(k)$ along $z$ axis which is

\begin{equation}\label{eq-ham-sys-3}
H(k)=-J_{mf1}\sum_{i,j}{\vec{S}_{i}\cdot\vec{S}_{j}}
-J_{mf2}\sum_{i,j}{\vec{S}_{i}\cdot\vec{S}_{j}},
\end{equation}
so that the energy of interface magnetoelectric interaction of the polarization at the site $k$ can be written as
\begin{equation}
\label{eq-ham-sysme-2} H_{mf}(k)=H(k)\ P_{k},
\end{equation}
In this expression $J_{mf1}$ is the
interaction parameter between the electric polarization component
$P_{k}$  at the interface ferroelectric layer and its NN spin on the
adjacent magnetic layer.
$J_{mf2}$ is the interaction parameter
between the electric polarization component $P_{k}$  at the interface
ferroelectric layer and the next NN spin on the
adjacent magnetic layer.  These interactions are shown
schematically in the Fig.~\ref{ref-fig2}.

%Fig2
\begin{figure}[h]
\begin{center}
\includegraphics[scale=0.33]{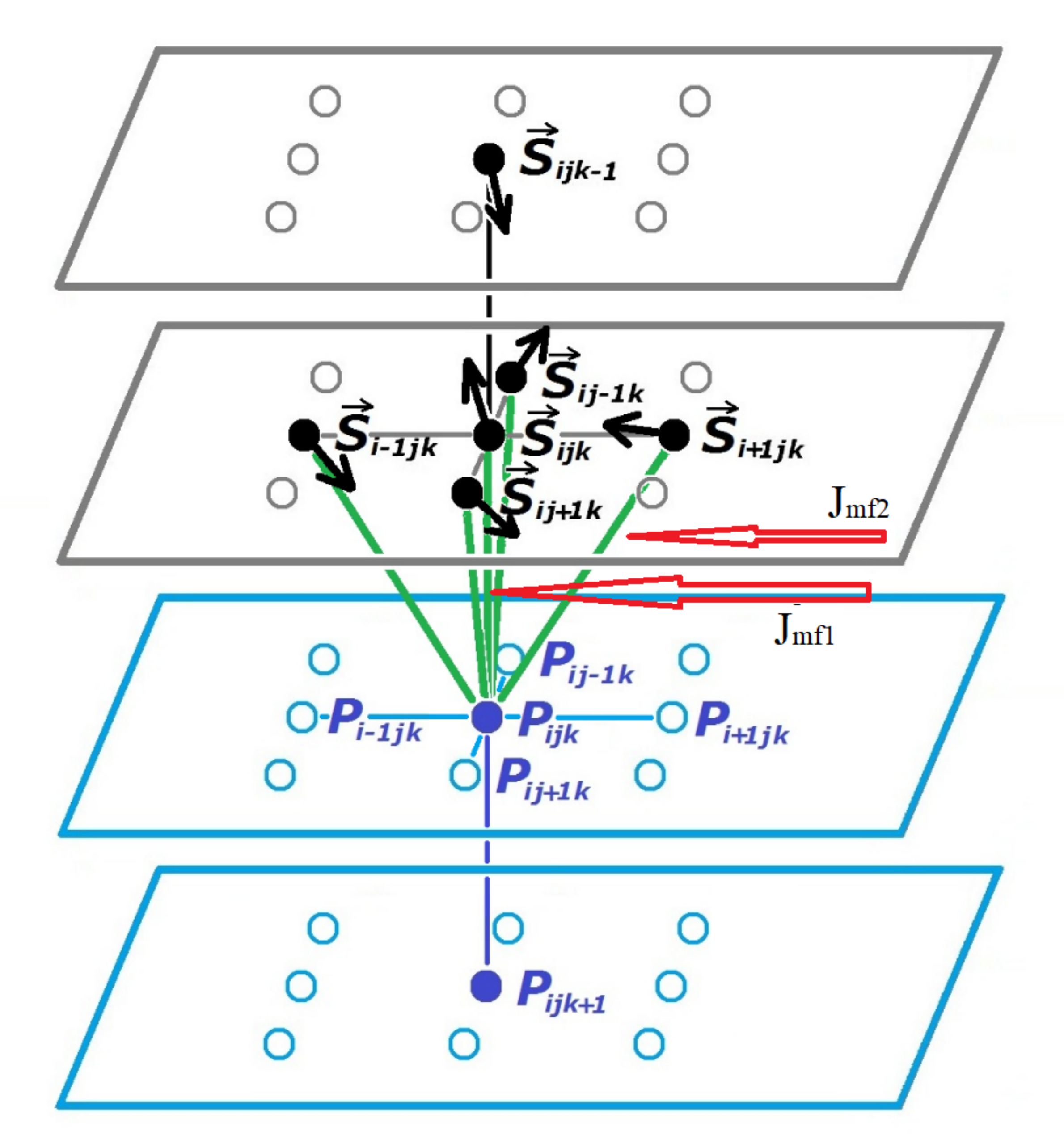}
\caption{Schematic representation of magnetoelectric interactions
at the interface between magnetic and ferroelectric layers.}\label{ref-fig2}
\end{center}
\vspace{-10pt}
\end{figure}

Note that the interface coupling described by Eq. (\ref{eq-ham-sysme-2}) is a scalar spin field acting on an electric polarization.  Later, in section \ref{other} we will suppose another form for the coupling: a scalar polarization field acting on the $z$ spin component.

\subsection{Ground state}

Let us take positive $J^{m}_{ij}$ and $J^{f}_{ij}$ so that magnetic layers are ferromagnetic and ferroelectric layers have parallel polarizations, in the ground state.

The relative orientation between two adjacent magnetic and ferroelectric layers depends on the signs of $J_{mf1}$ and $J_{mf2}$. There are two simple cases:

i) if they are both positive, then spins and polarizations are parallel in the ground state (GS)

ii) if they are negative, then spins are antiparallel to polarizations in the GS.

The complicated case occurs when $J_{mf1}$ and $J_{mf2}$ have opposite signs. In this case, there is a competition between them which gives rise to some degree of frustration. For example, when $J_{mf1}>0$ and $J_{mf2}<0$  we have the situation where NN interaction wants $\vec S$ and $\vec P$ to be parallel, while the NNN interaction wants them to be antiparallel. Depending on their respective amplitudes, one configuration wins over the others.

Let us write the GS energy of a spin at the interface in zero fields
\begin{equation}
E_{1}=-Z_{1}J^{m}-Z_{2}J_{mf1}-Z_{3}J_{mf2}
\end{equation}
where the coordination numbers are  $Z_{1}=4, Z_{2}=1, Z_{3}= 4$ for a simple cubic
lattice.

%Fig3
\begin{figure}[h]
\begin{center}
\includegraphics[scale=0.45]{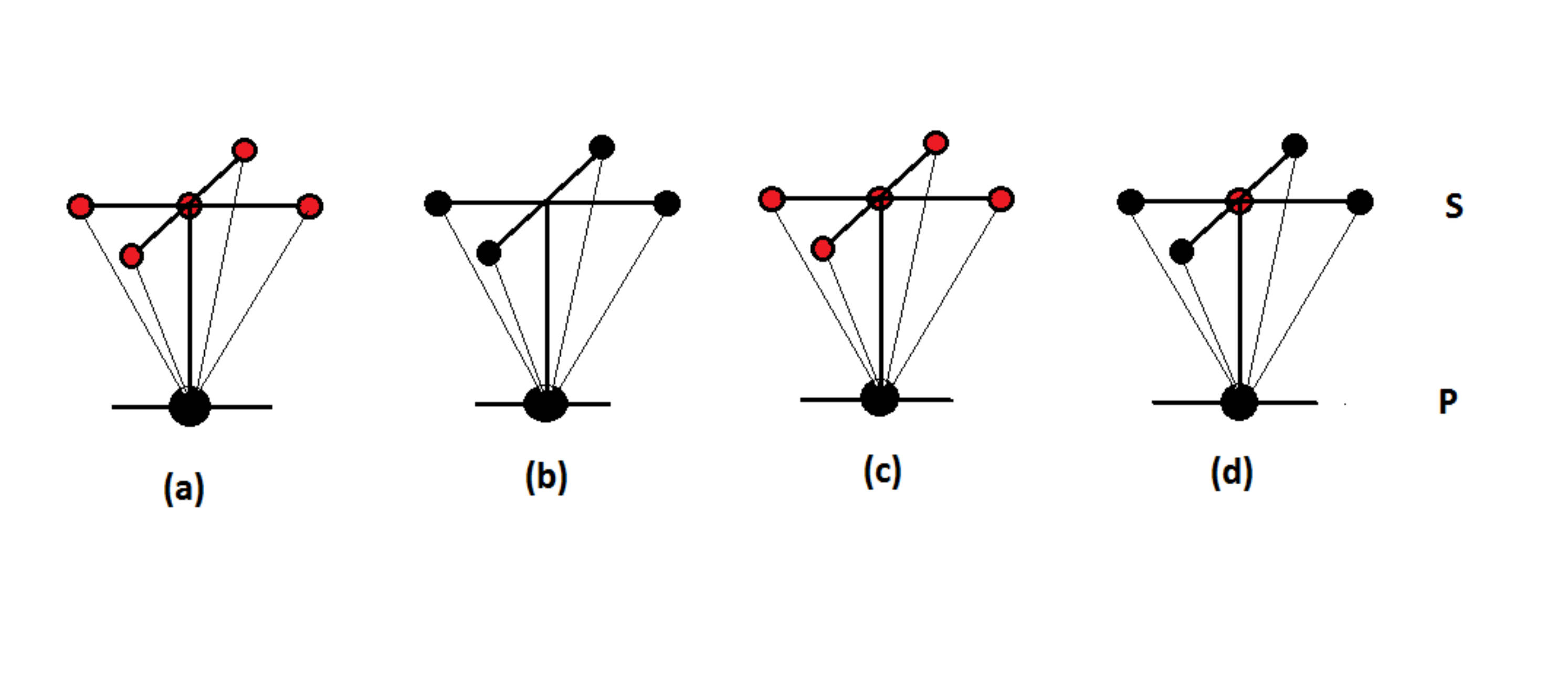}
\caption{Ground state spin configurations with energies from $E_1$ to $E_4$ (a to d, respectively) depending on the interface interactions between magnetic and ferroelectric layers.  Lower black circles are $P=1$ (up), upper black circles are up spins, red circles are down spins.}\label{GS}
\end{center}
\vspace{-10pt}
\end{figure}
For $J_{mf1}<0$, the four spin configurations (see Fig. \ref{GS})
\begin{eqnarray}
E_{1}&=&-Z_{1}J^{m}-Z_{2}|J_{mf1}|-Z_{3}|J_{mf2}|\label{eq-gs}\\
E_{2}&=&-Z_{1}J^{m}+Z_{2}|J_{mf1}|-Z_{3}J_{mf2}\label{eq-gs2}\\
E_{3}&=&-Z_{1}J^{m}-Z_{2}|J_{mf1}|+Z_{3}J_{mf2}\label{eq-gs3}\\
E_{4}&=&+Z_{1}J^{m}-Z_{2}|J_{mf1}|-Z_{3}J_{mf2}\label{eq-gs4}
\end{eqnarray}
where $E_1$ is the energy of the state where all spins are down, all polarizations are up with $J_{mf2}<0$ (Fig. \ref{GS}a). Other energies correspond to the spin configurations shown in Fig. \ref{GS}: $E_2$ to Fig. \ref{GS}b with $J_{mf2}>0$, $E_3$ to Fig. \ref{GS}c with $J_{mf2}>0$ and $E_4$ to Fig. \ref{GS}d with $J_{mf2}>0$.

The system will choose the GS depending on the values of $J_{mf1}$ and $J_{mf2}$. For simplicity, let us confine ourselves to GS configurations where all magnetic spins are parallel and all polarizations are parallel, i. e. the first three configurations. This choice is possible if the intralayer interactions $J^m>0$ and $J^f>0$ are sufficiently strong.

The state $E_1$ is chosen if

\begin{equation}\label{eq-gs1}
E_{1}<E_{2}, E_{1}<E_{3}, E_{1}<E_{4}
\end{equation}
Solving these inequalities we have
\begin{equation}\label{eq-gs1a}
 J_{mf2} <0, |J_{mf1}|<\frac{Z_{1}J^{m}-Z_{3}J_{mf2}}{Z_{2}},
 |J_{mf1}|>0
\end{equation}
namely,
\begin{equation}\label{eq-gs1b}
 J_{mf2} <0, |J_{mf1}|<4(J^{m}-J_{mf2}),J_{mf1}<0
\end{equation}

Now we
suppose $J_{mf2}>0$ then the GS will change to $E_{2}$. The critical value of
$J_{mf2}$ and $J_{mf1}$ are determined by solving
\begin{equation}\label{eq-gs2a}
E_{2}<E_{1},E_{2}<E_{3},E_{2}<E_{4}
\end{equation}
We have
\begin{equation}\label{eq-gs2b}
J_{mf2} >0, J_{mf1}<\frac{Z_{3}J^{mf2}}{Z_{2}},
|J_{mf1}|<\frac{Z_{1}J^{m}}{Z_{2}}
\end{equation}
namely,
\begin{equation}\label{eq-gs2c}
J_{mf2} >1/4|J_{mf1}|,J_{mf2} >0, |J_{mf1}|<4 J_{m}
\end{equation}

The GS is $E_3$ if we have
\begin{equation}\label{eq-gs3a}
E_{3}<E_{1},E_{3}<E_{4}, E_{3}<E_{2}
\end{equation}
We get
\begin{equation}\label{eq-gs3b}
|J_{mf1}| >0, J_{mf2}<\frac{Z_{2}|J_{mf1}|}{Z_{3}}, J_{mf2}<\frac{Z_{1}J^{m}}{Z_{3}}
\end{equation}
or
\begin{equation}\label{eq-gs3c}
J_{mf2} <1/4|J_{mf1}|,J_{mf2} <J^{m}, |J_{mf1}|>0
\end{equation}

In MC simulations shown below, care should be taken to choose the right GS according to values and signs of the interface interactions to avoid metastable states at low temperatures ($T$).

Note that we have taken $J_{mf1}<0$ in the above spin configurations. This is intended to have spins antiparallel to the magnetic field applied in the $+z$ direction so as to have a phase transition at a temperature with a finite field.

\section{Monte Carlo Simulation}\label{sect-results}
For MC simulations we use the Metropolis
algorithm and a sample size $L\times L\times L_{z}$ with $L =
40,60,80, 100$ for detection of lateral size effects and $L_{z} = 8,
16,12, 24$ for thickness effects. When we investigate the effects of
the magnetoelectric coupling on the magnetic, ferroelectric and
interface properties, for simplicity we take the same size and thickness for the
ferroelectric and magnetic layers (for example if $L_{z}=8$ - it means $L_z^m=4$
magnetic layers and $L_z^f=4$ ferroelectric layers). Exchange parameters between
intralayer spins and intralayer polarizations are taken to be $J^{m}=J^{f}=1$ for the
simulation.

For MC simulation we perform the cooling from the disordered phase: electrical
polarizations of $\pm{1}$ are randomly assigned at lattice sites
in the ferroelectric layers, in the $z$ direction. In the ferromagnetic layers spins with
$|\vec{S}|=1$ are also randomly assigned in any direction,
following in the spatial uniform distribution. At each $T$, new
random $\vec{S_{i}}$ and $P_{i}$ were chosen, and the energy
difference caused by this change is calculated. This change is
accepted or rejected according to the Metropolis algorithm. In order
to ensure the convergence of the observables, the
lattice is swept 100000 times, where each time is considered as one
MC step (MCS) that can be taken as the time scale of
simulations. The observables of interest such as the averages of layer electric polarizations
$P$,  and layer magnetization $M$, are calculated over
the following 50000 MCS. These quantities are defined as

\begin{equation}\label{eq-orpar1}
P(n)=\frac{1}{L^{2}}\langle{|\sum_{i\in n} P_{i}|}\rangle
\end{equation}
\begin{equation}\label{eq-orpar2}
M(m)=\frac{1}{L^{2}}\langle{|\sum_{j\in m}\vec S_{j}}|\rangle
\end{equation}
where $\langle...\rangle$ is the time average, and the sums on  $i$ and $j$ are performed over the lattice sites belonging to the ferroelectric layer $n$ and the lattice sites belonging to the magnetic layer $m$, respectively.
The process is repeated for a lower $T$ down to the desirable lowest one.
We also perform  the heating, starting from the GS spin configuration.

\subsection{Zero fields}
Monte Carlo results for energy, magnetization and polarization and their susceptibilities obtained by heating the system from
the initial spin configuration of GS energy $E_1$ are shown
in Fig.~\ref{ref-fig2}. Of course, starting from different initial spin configurations which are not the GS will lead to the same thermodynamic equilibrium but the equilibrating time is longer in particular at low $T$.

%Fig4
\begin{figure}[h]
\vspace{10pt}
\begin{center}
\includegraphics[scale=0.36]{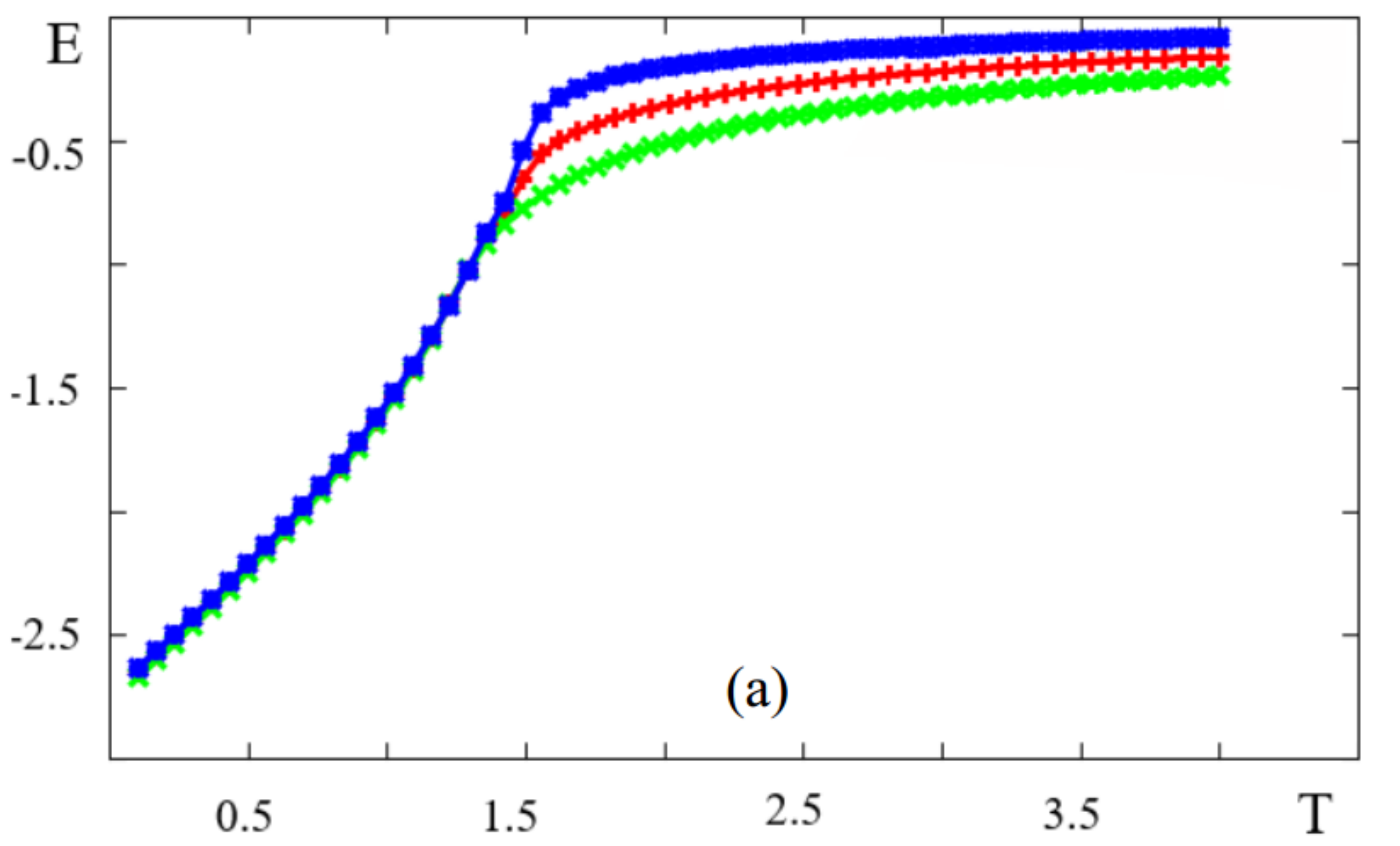}
\includegraphics[scale=0.36]{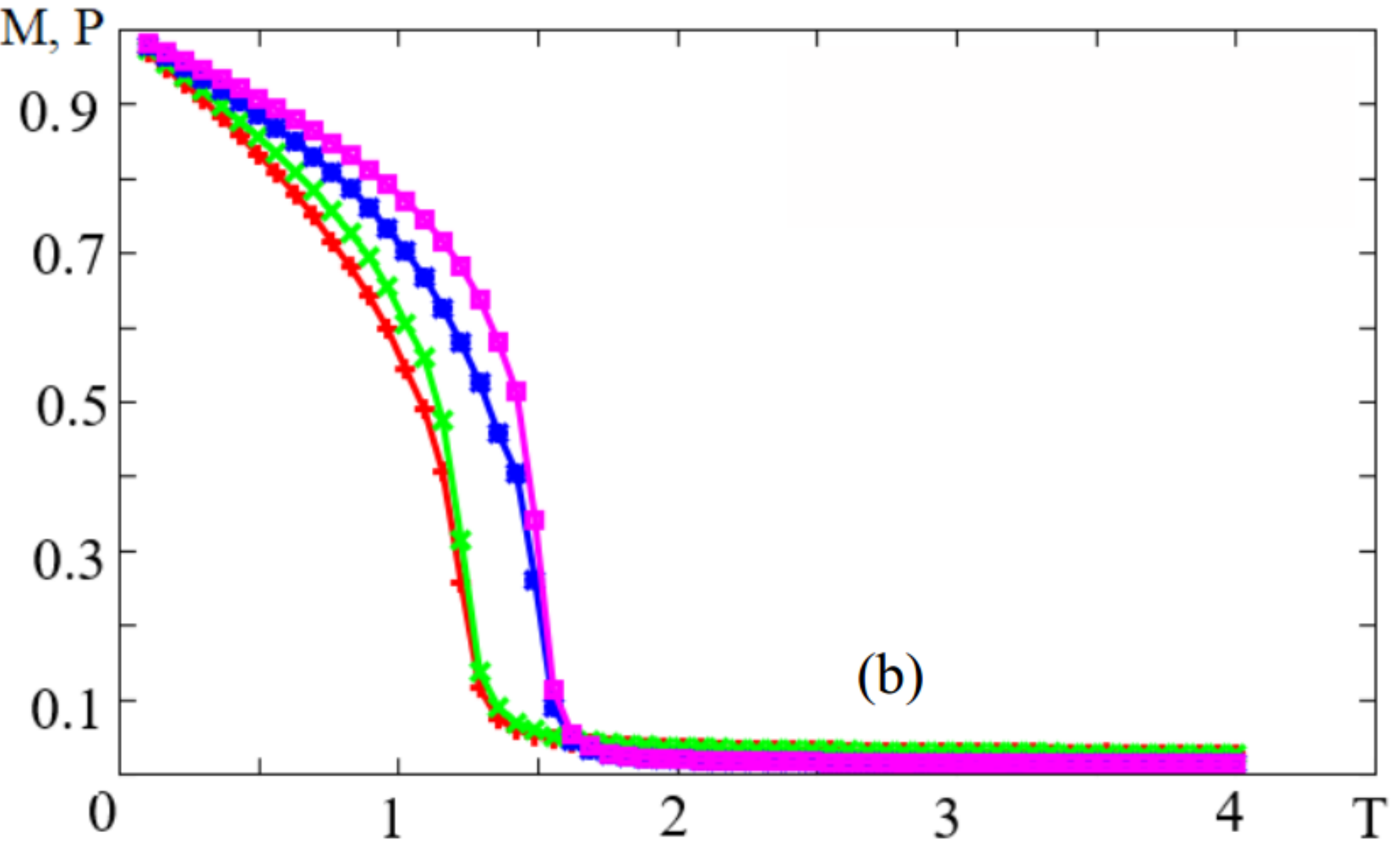}
\includegraphics[scale=0.36]{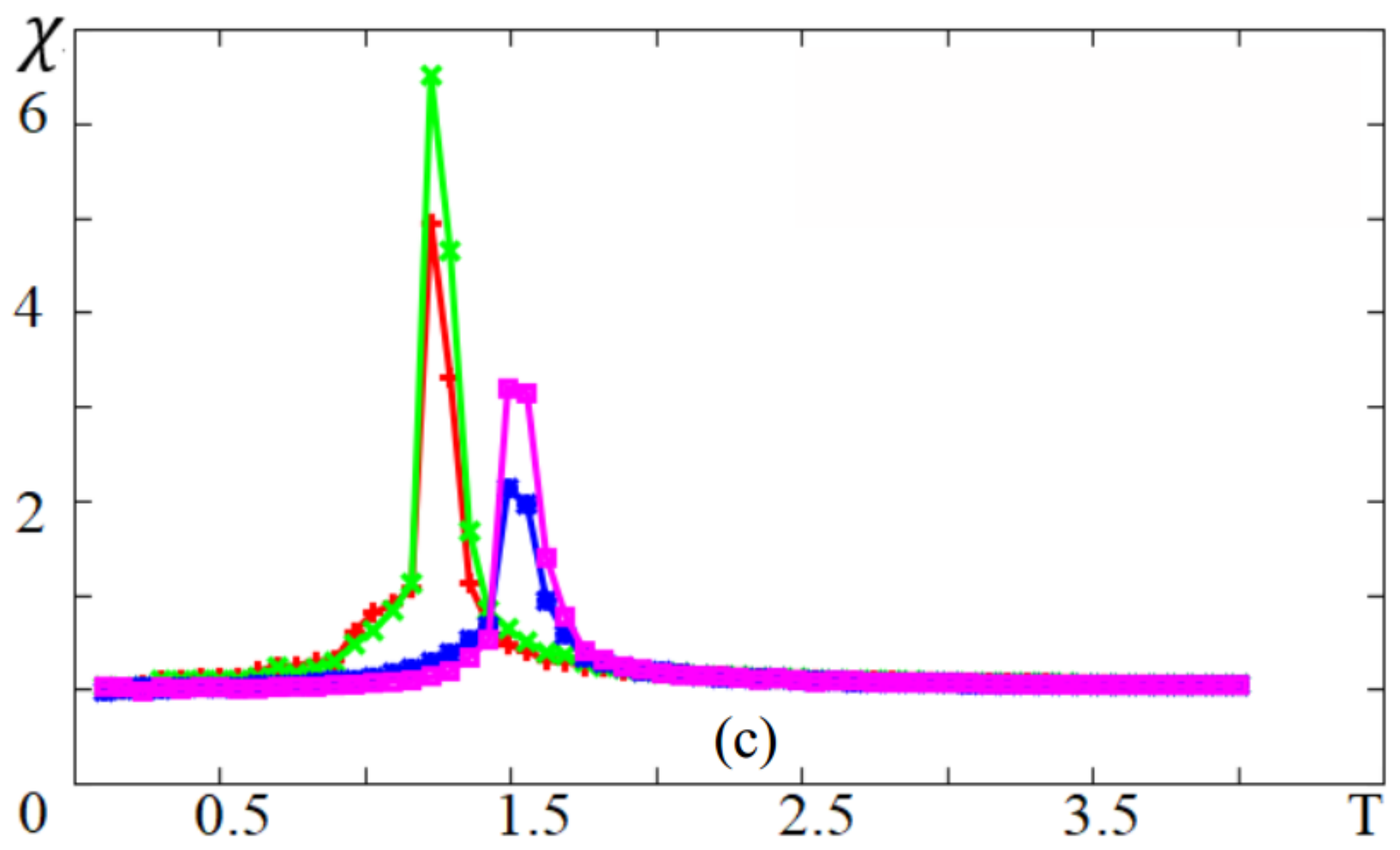}
\end{center}
\vspace{10pt} \caption{(a) Energy versus $T$. Red line:
energy for total superlattice, green line: energy of magnetic
layers, blue line: energy of ferroelectric layers ; (b)
Magnetization and polarization versus $T$. Red line:
magnetization of interface magnetic layers, green line: magnetization
of interior magnetic layers, blue line: polarization of interface
ferroelectric layers, purple line: polarization of interface
ferroelectric layers ; (c) Susceptibilities versus $T$, with the same color code.
$J^{m}=1,
J_{mf1}=-0.15, J_{mf2}=-0.135$ corresponding to the GS with energy
$E_{1}$.} \label{ref-fig2} \vspace{10pt}
\end{figure}

\begin{figure}[h]
\vspace{10pt}
\begin{center}
\includegraphics[scale=0.36]{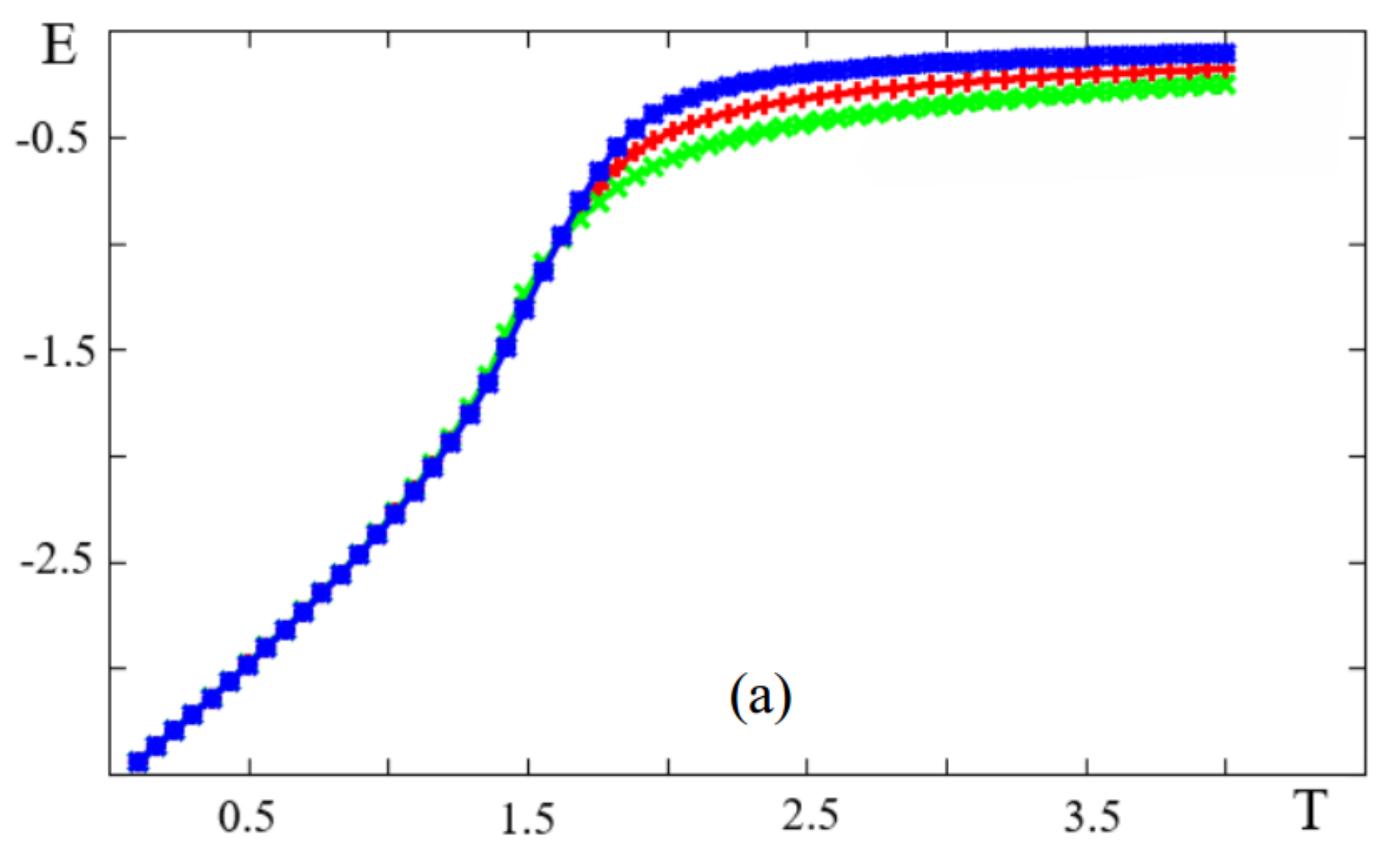}
\includegraphics[scale=0.36]{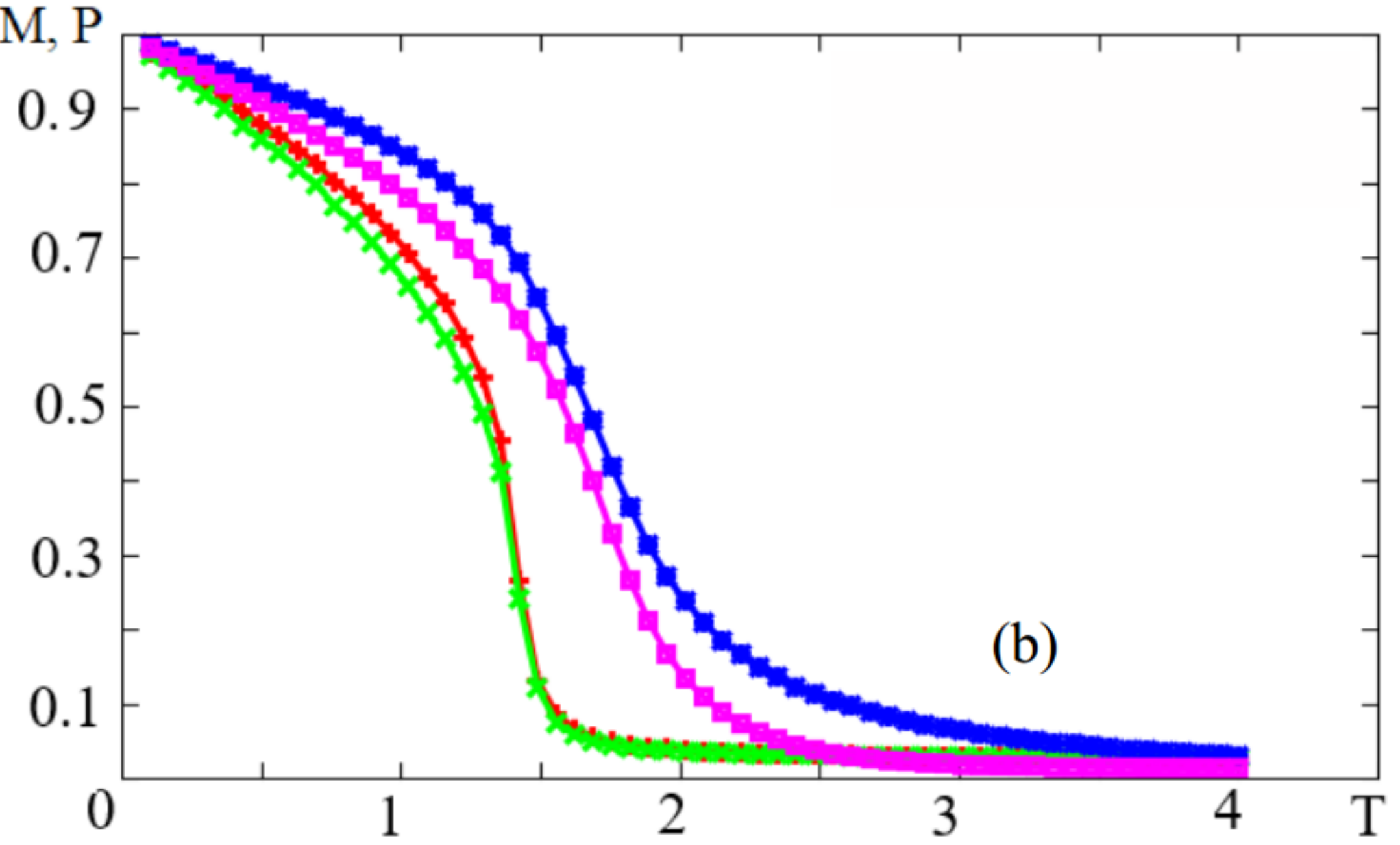}
\includegraphics[scale=0.40]{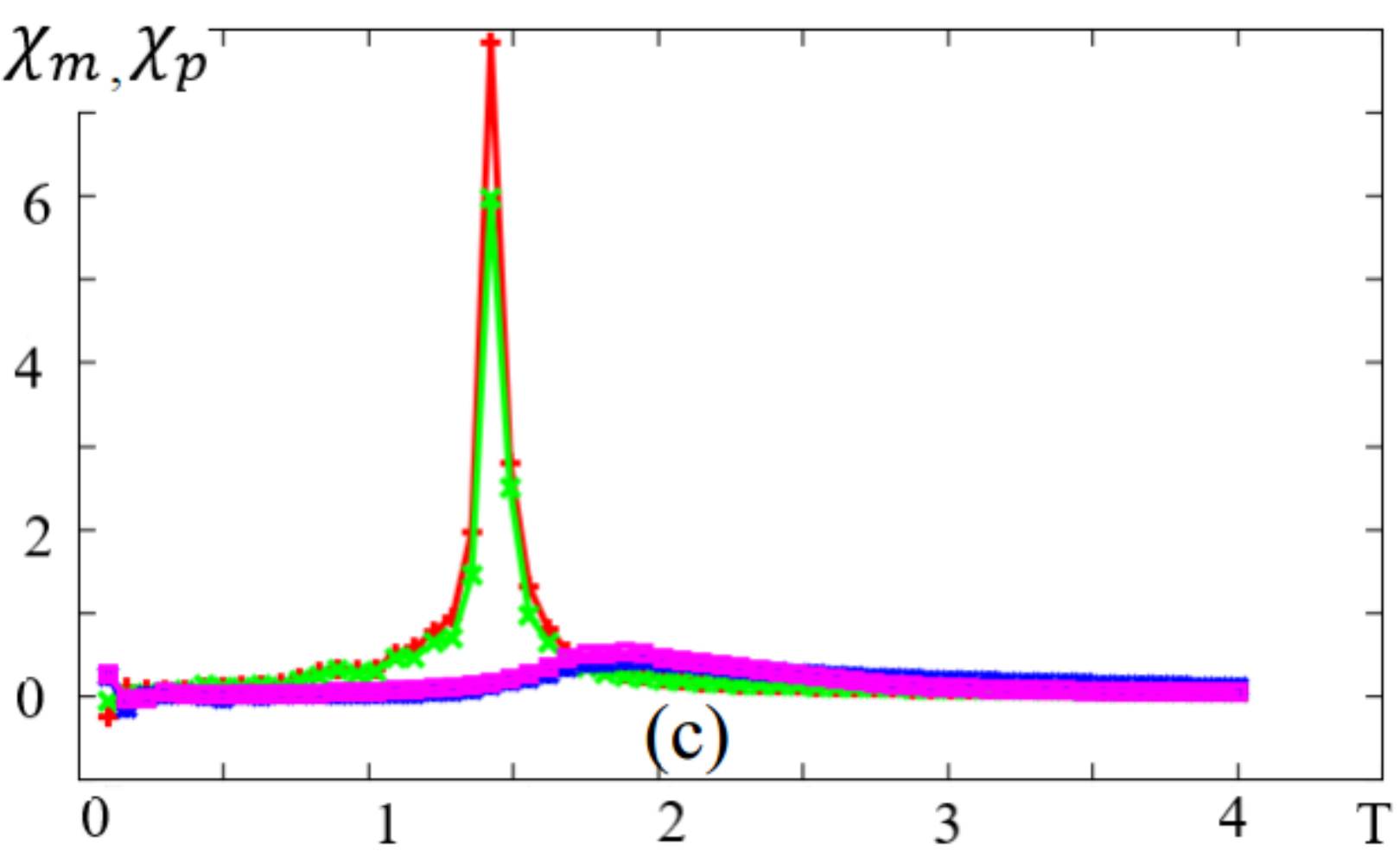}
\end{center}
\vspace{-20pt} \caption{(a) Energy versus $T$. Red line:
energy for total superlattice, green line: energy of magnetic
layers, blue line: energy of ferroelectric layers ; (b)
Magnetization and polarization versus $T$. Red line:
magnetization of interface magnetic layers, green line: magnetization
of interior magnetic layers, blue line: polarization of interface
ferroelectric layers, purple line: polarization of interface
ferroelectric layers ; (c) Susceptibilities versus $T$, with the same color code.    $J^{m}=1,
J_{mf1}=-0.15,J_{mf2}=0.8$ corresponding to the GS with energy
$E_{2}$.  }\label{ref-fig2c} \vspace{20pt}
\end{figure}

MC results for energy, magnetization and polarization and their susceptibilities obtained by heating the
initial spin configurations of energy $E_2$ are shown in
Fig.~\ref{ref-fig2c}. For $E_3$, the results are qualitatively similar (not shown). respectively.

The above figures show that the energy and other physical quantities well behave at low $T$ (no metastability) if we choose the correct GS according to the interface interaction.  Note that the ferroelectric films undergo a phase transition at a temperature higher than that of the magnetic films.
This is due to the Ising-like nature of the ferroelectric polarizations (in the bulk, the transition temperature is inversely proportional to $N$, the spin components, $\sim 1/N$).
Also, the interface layers have lightly smaller order parameter than those inside the films.

%Fig5

%Fig6
%\begin{figure}[h]
%\vspace{10pt}
%\begin{center}
%\includegraphics[scale=0.36]{fig2_3.pdf}
%\includegraphics[scale=0.36]{fig2_4.pdf}
%\includegraphics[scale=0.36]{fig2_5.pdf}
%\end{center}
%\vspace{20pt} \caption{(a) Energy versus $T$. Red line:
%energy for total superlattice, green line: energy of magnetic
%layers, blue line: energy of ferroelectric layers ; (b)
%Magnetization and polarization versus $T$. Red line:
%magnetization of interface magnetic layers, green line: magnetization
%of interior magnetic layers, blue line: polarization of interface
%ferroelectric layers, purple line: polarization of interface
%ferroelectric layers ; (c) Susceptibilities versus $T$, with the same color code.
% $ J^{m}=1,
%J_{mf1}=-0.15,J_{mf2}=0.02$ corresponding to the GS with energy
%$E_{3}$. }\label{ref-fig2_3} \vspace{20pt}
%\end{figure}
%

\subsection{Particular Case $J_{mf1}=J_{mf2}$}

In this section we present the results of the MC simulations in the particular case where $J_{mf1}=J_{mf2}$.  We will compare these results with results from the MF theory.

\begin{figure}
\begin{center}
\includegraphics[scale=0.46]{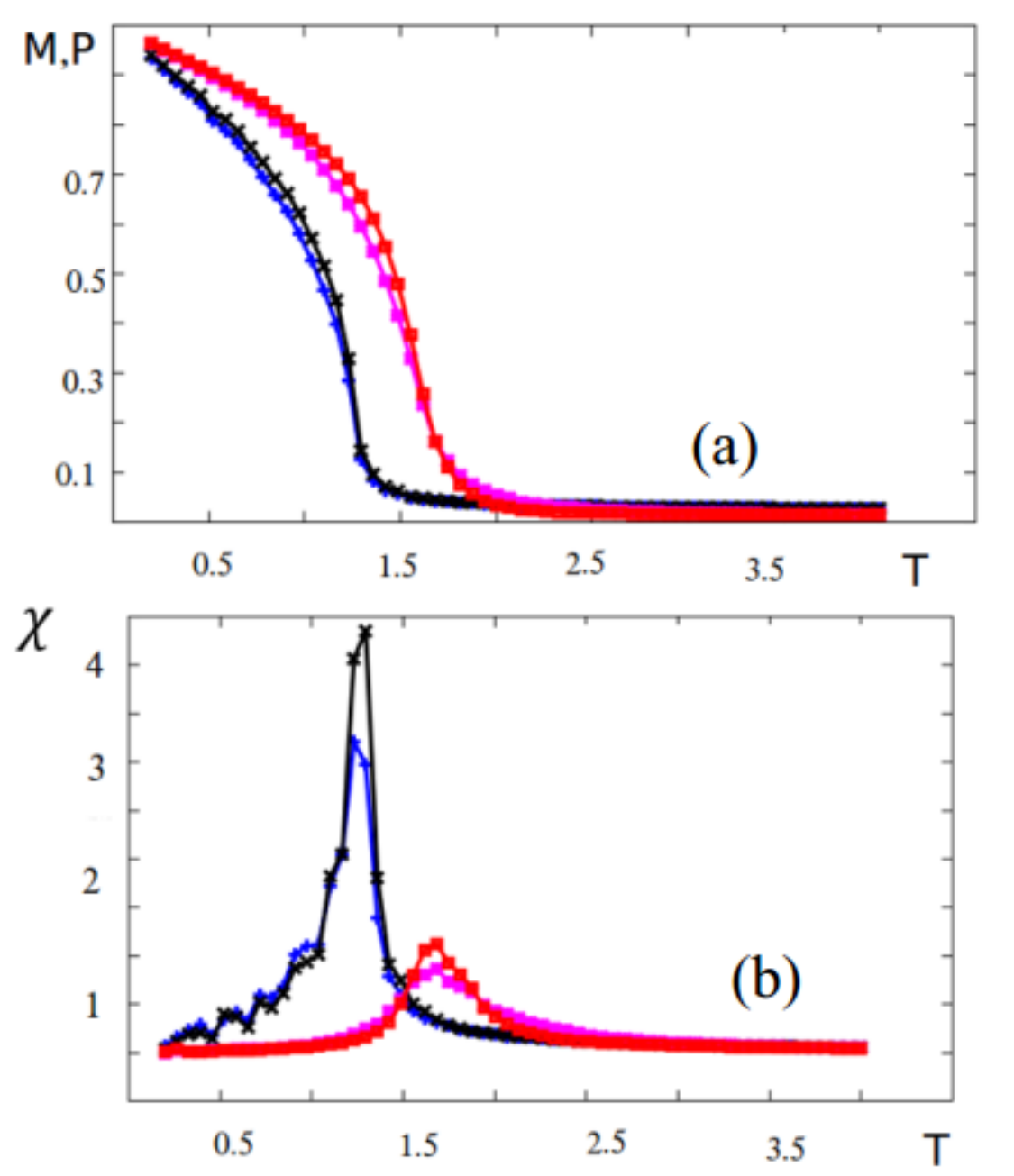}
\includegraphics[scale=0.46]{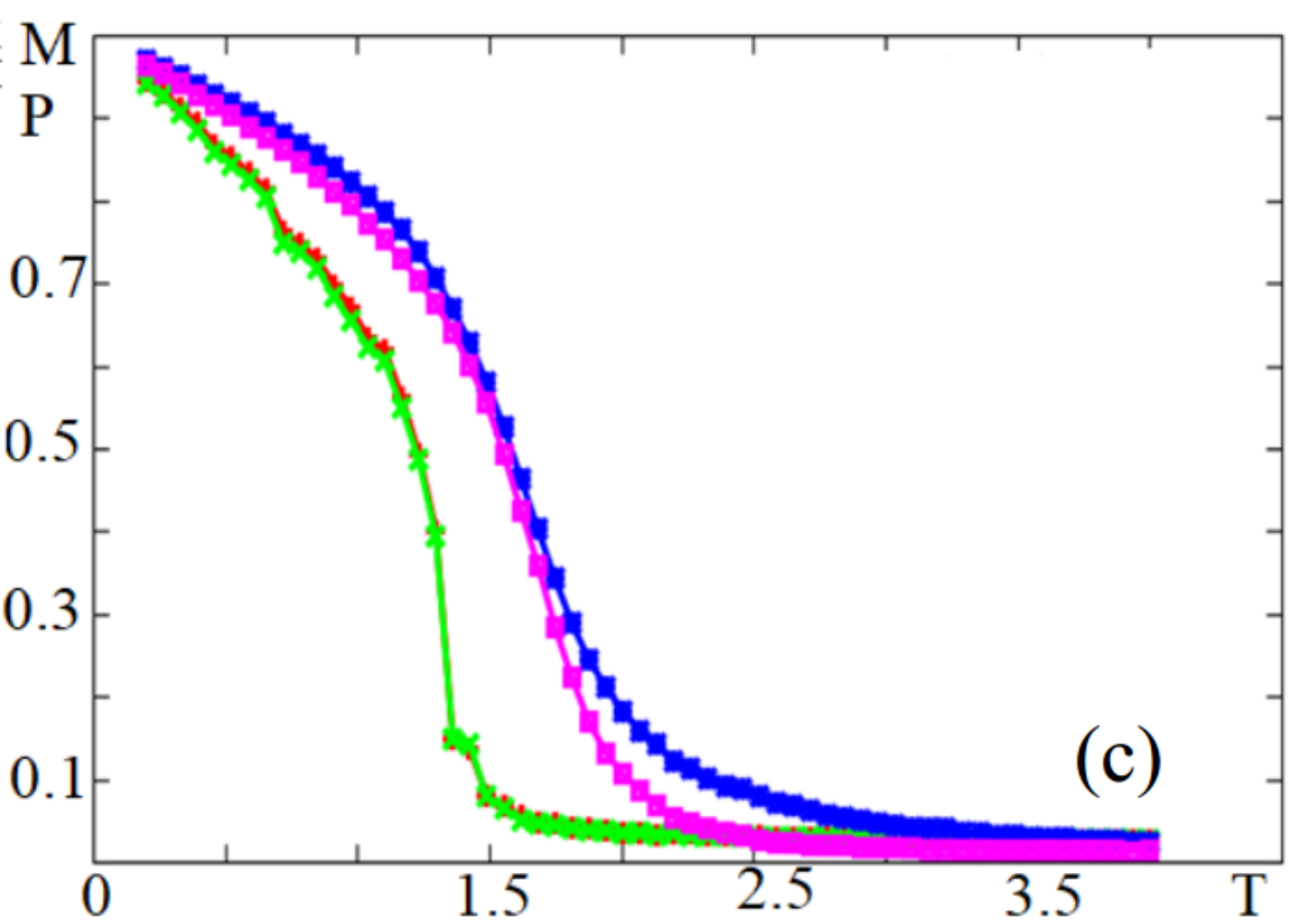}
\end{center}
\vspace{15pt} \caption{(a) Temperature dependence of layer magnetizations and
layer polarizations, (b) layer susceptibilities, in the case $J^{m}=1$, $J^{f}=1$,
$J_{mf}=-0.15$, $H^z=E^z=0$, $L=40$, $L_{z}=8$. Blue
squares for the first layer and fourth magnetic layers (interface layers), black
circles for the second and third (interior magnetic layers), magenta squares for the first and
fourth (interface) ferroelectric layers, red for the second and third
interior ferroelectric layers, respectively. $J_{mf1}=J_{mf2}$;
(c) Temperature dependence of layer magnetizations and polarizations for $J^{m}=1$, $J^{f}=1$, $J_{mf}=-0.55$,
$H^z=E^z=0$, $|\vec{S}|=1$, $P=\pm 1$. The coupling used is Eq. (\ref{eq-ham-sys-2}). Red and green lines are for magnetic interface and inner layers, magenta (blue) line for the interface (inner) ferroelectric layer. \label{ref-fig4}} \vspace{15pt}
\end{figure}

Results for the temperature dependence of layer magnetizations, polarizations, and susceptibilities of the system are shown in Fig.~\ref{ref-fig4} for
$J_{m}=J_{f}=1, J_{mf}=J_{mf1}=J_{mf2}=-0.15, -0.55$. For $J_{mf}$, these curves present a sharp second-order
transitions at $T_{c}^m\simeq 1.32$ for magnetic layers and
$T_{c}^f\simeq 1.84$ for ferroelectric layers.

\begin{figure}[h]
\vspace{10pt}
\begin{center}
\includegraphics[scale=0.46]{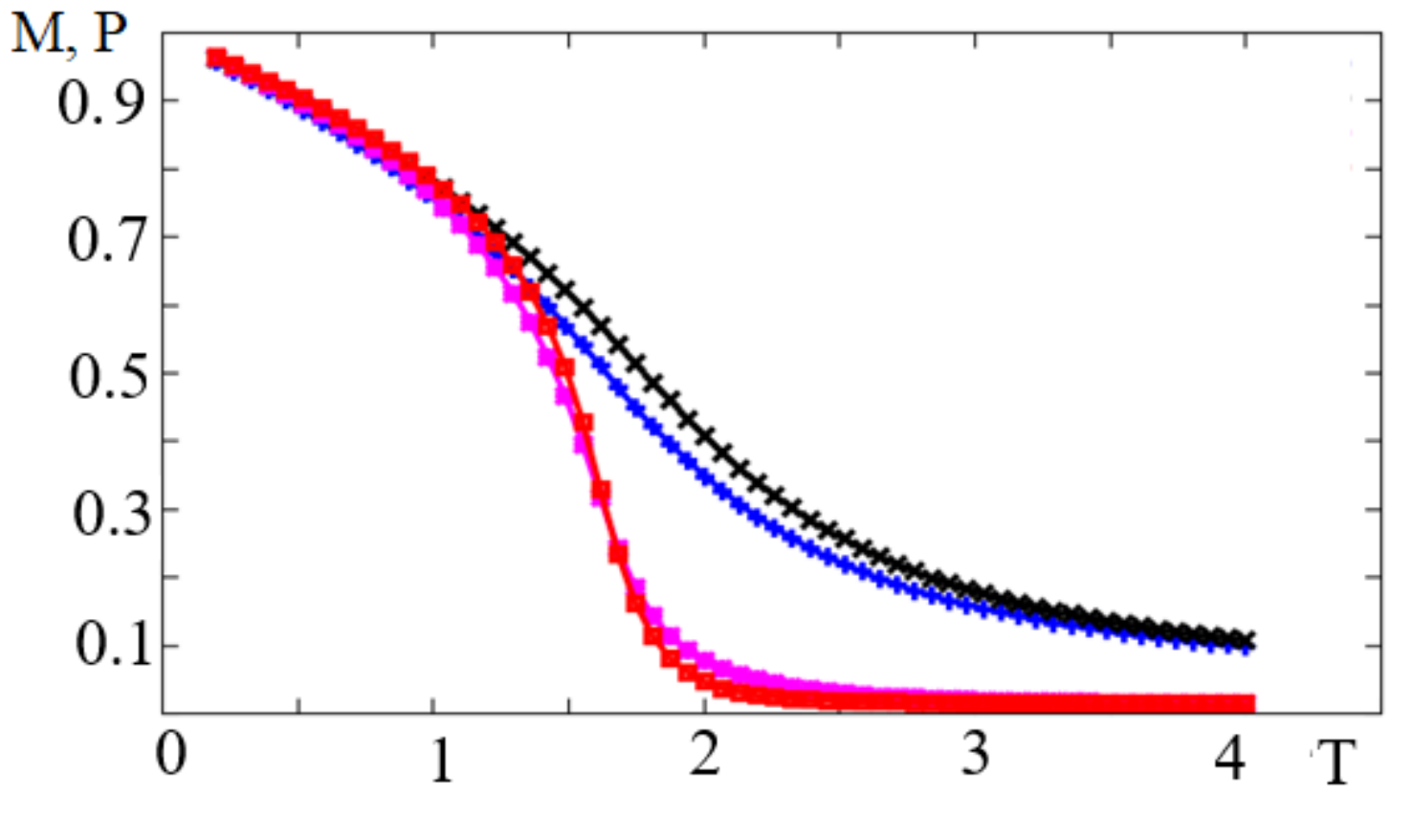}
\end{center}
\vspace{20pt} \caption{Temperature dependence of layer magnetizations and
layer polarizations in case $J^{m}=1$, $J^{f}=1$,
$J_{mf}(=J_{mf1}=J_{mf2})=-0.15$, $H^z=0.7$ and $E^z=0$, $L=40$, $L_{z}=8$. Blue
squares for the first layer and fourth magnetic layers (interface layers), black
circles for the second and third (interior magnetic layers), magenta squares for the first and
fourth (interface) ferroelectric layers, red for the second and third
interior ferroelectric layers, respectively.
\label{ref-fig5}} \vspace{20pt}
\end{figure}

The results with an external magnetic field $H^z=0.7$ are
shown in Fig.~\ref{ref-fig5} for the order
parameters. We can see that in this case the magnetic subsystem does not
undergo a phase transition as a ferromagnet in a field. On the contrary, there is a second-order
transition for ferroelectric layers at $T_{c}\simeq 1.84$.

%Fig6

%Fig7

\begin{figure}[h]
\vspace{10pt}
\begin{center}
\includegraphics[scale=0.46]{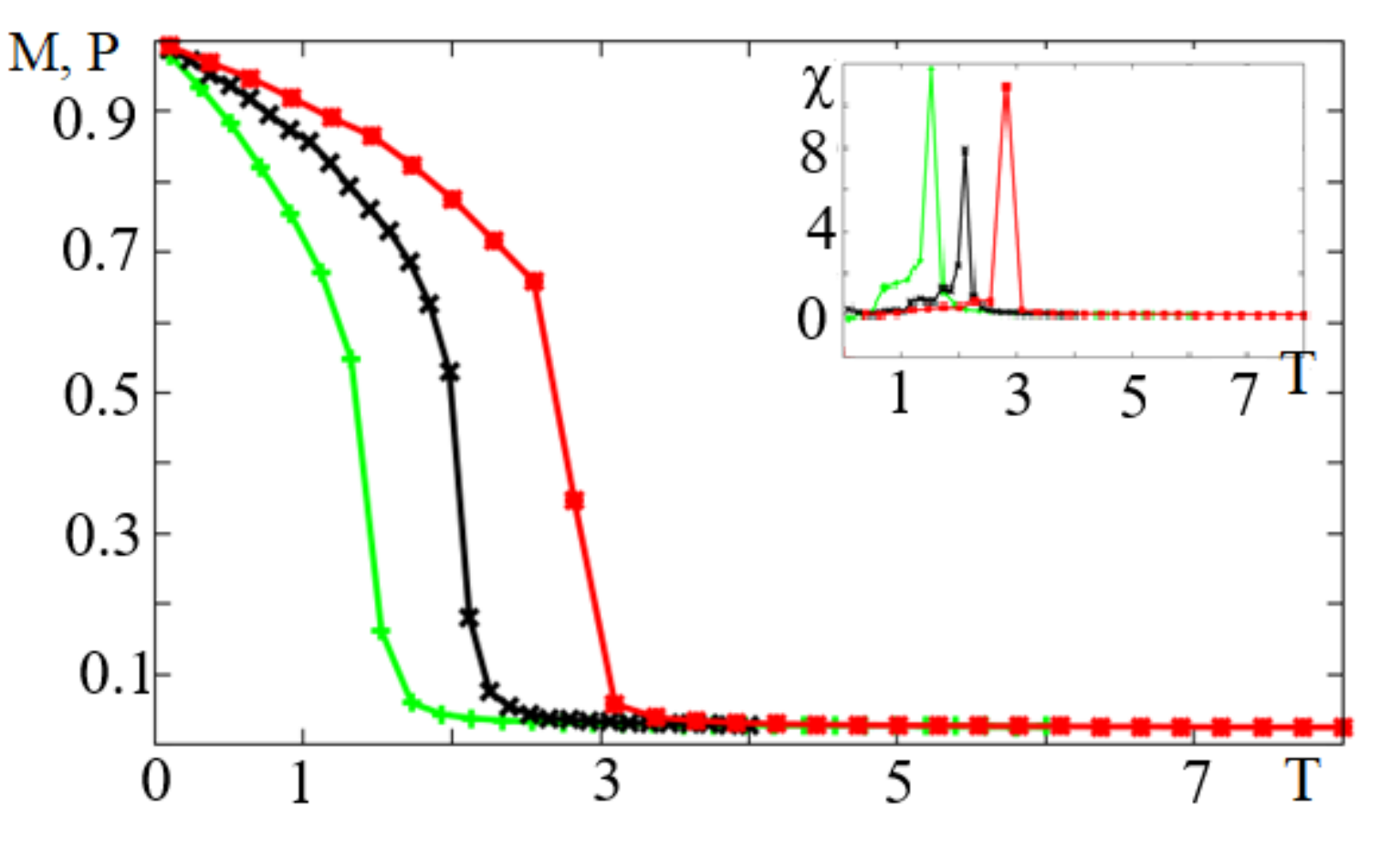}
\end{center}
\vspace{-20pt} \caption{Temperature dependence of the total magnetization and
susceptibility in case $J^{m}=1$, $J^{f}=1$,
$J_{mf}=-1,-3,-5.5$, $H^z=E^z=0$, $L=40$, $L_{z}=8$. Red
lines for $J_{mf}=-5.5$, black lines for $J_{mf}=-3$, green lines for
$J_{mf}=-1$. \label{ref-fig7}
}\vspace{20pt}
\end{figure}

With increasing $J_{mf}$, the system undergoes
a first-order transitions. Fig.~\ref{ref-fig7} shows the total magnetization $M$
and susceptibility versus $T$ for several values of $J_{mf}$ in the
cross-over region from second to first order. The second-order phase
transitions starts at $J_{mf}=0$ with $T_{c} \approx 2.456$, it
decreases as $J_{mf}$ increases.

%Fig8

\begin{figure}
\vspace{10pt}
\begin{center}
\includegraphics[scale=0.60]{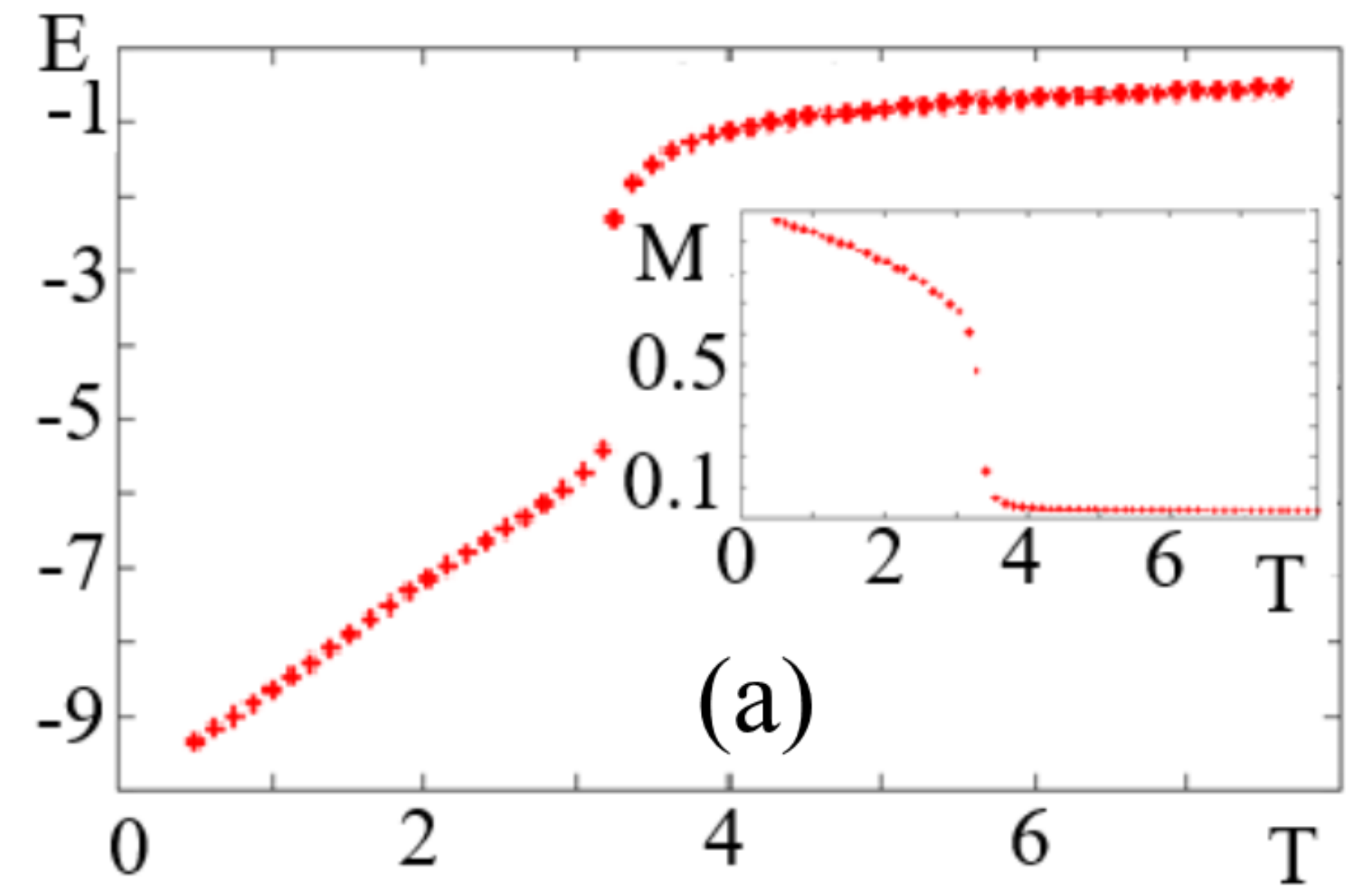}
\includegraphics[scale=0.42]{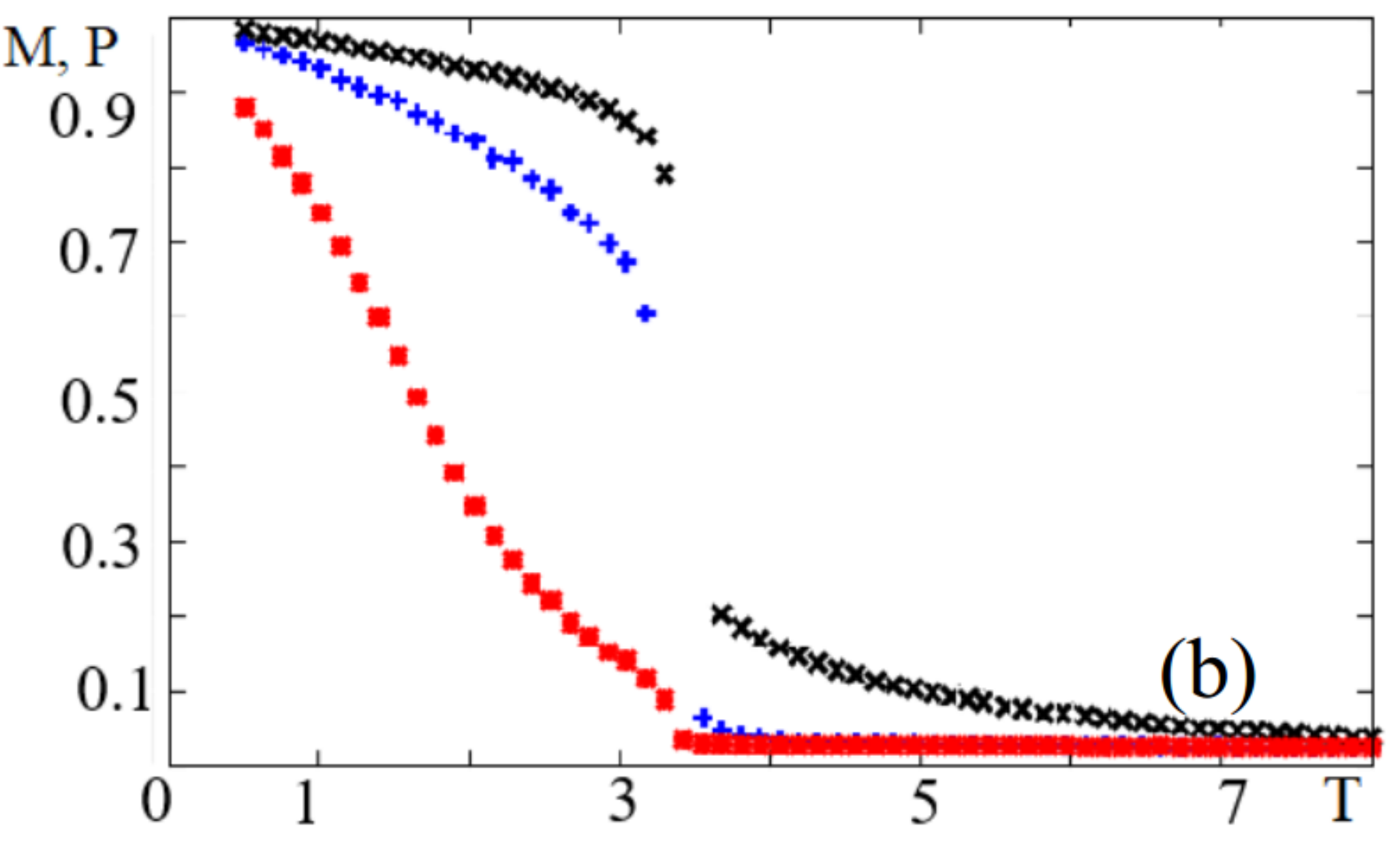}
\end{center}
\vspace{-20pt} \caption{(a) Energy, layer magnetization (inset)
versus $T$ ; (b) Interface layer magnetization (blue +), interior layer magnetization (red squares), interface polarization (black X) and interior layer polarization (magenta X overlapped under red squares),  versus
$T$.  $J^{m}=1$, $J^{f}=1$, $J_{mf}=-9.5$, $H^z=E^z=0$
$L=40$, $L_{z}=8$.\label{ref-fig8}}\vspace{10pt}
\end{figure}

We show in Fig.~\ref{ref-fig8} the case of $J_{mf} = -9.5$ where one
can observe a discontinuity at the transition temperature
$T_{c}\simeq 3.45$ for the interface  magnetic layer and
$T_{c}\simeq 3.49$ for the interface ferroelectric layer. Only
layer $1$ and $4$ for magnetic and ferroelectric systems have a
phase transition.  Their order parameters strongly fall down at the
transition temperature. This result is confirmed by several
independent simulations.  We calculate the transition temperature as a function of
$J_{mf}$. We keep $J_{mf}$ constant, change the temperature and we
take the transition temperature at the peak of the magnetic and
ferroelectric susceptibility $\chi$.  
Note that for the strong interface coupling $J_{mf}=-9.5$, the interface order (black and blue curves in Fig. \ref{ref-fig8}b)  is so strong that it acts on the interior layer as an external field which does not allow the interior layer order parameter to go to zero: as a consequence,  the interior layer undergoes only a smooth change of curvature at $T\simeq 1.5$  and  falls to zero with the interface magnetic layer at $T_{c}\simeq 3.45$ (see red curve in Fig. \ref{ref-fig8}b).

%Fig9

%Fig10
\begin{figure}[h]
\vspace{-10pt}
\begin{center}
\includegraphics[scale=0.45]{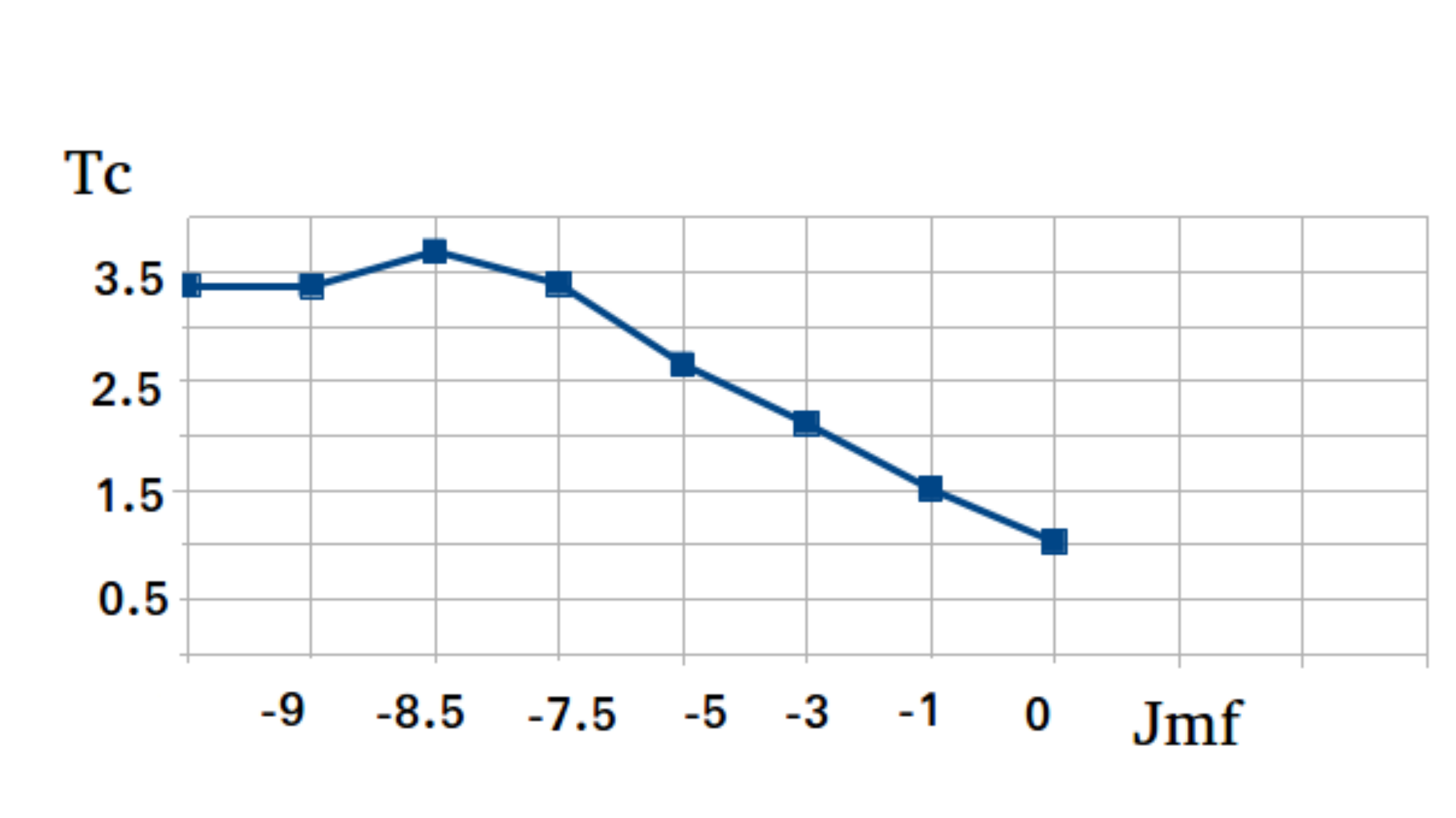}
\end{center}
\vspace{-20pt} \caption {Phase diagram in $T-J_{mf}$ plane. Here
$J^{m}=1$, $J^{f} = 1$, $H^z=E^z=0$.\label{ref-fig10}}
\vspace{20pt}
\end{figure}

The results for the transition temperature $T_c$ are shown in Fig.~\ref{ref-fig10} as a function  of $J_{mf}$. One can see that the transition temperature increases when
we increase the values of $|J_{mf}|$. $T_c$ has a maximum at $J_{mf}=-8.5$.   The second-order phase transition
starts at $J_{mf}=0$ and becomes a first-order phase transition below $J_{mf}=-9$
(see Fig.~\ref{ref-fig8} for $J_{mf}=-9.5$).

\section{Another model of interface interaction}\label{other}
Let us show
some results for the another model of magnetoelectric interaction given in the
form
\begin{equation}\label{ham-sysme-3}
H^{1}_{mf}=-J_{mf}\sum_{i,j,k} P_{i} P_{j} S_{i}^z
\end{equation}

We show in Fig.~\ref{ref-fig14} layer magnetizations and polarizations as functions
of interface coupling $J_{mf}$
For small values of
the magnetoelectric interaction, magnetic and ferroelectric
layers undergo phase transitions of different orders:
magnetic layers
undergo a phase transition of the first order, while ferroelectric layers undergo a
the second-order transition, at temperatures $T_{c}\approx 0.63$
and $T_c\approx 1.52$, respectively.

With an increase of the magnetoelectric interaction
$J_{mf}$ between the magnetic and ferroelectric subsystems, an
unusual phenomenon is observed: the interface layers after $J_{mf}=-3.5$
undergo phase transitions of the first order. Note that in the model
considered at the beginning of this article this occurs at large
values of $J_{mf}=-9.5$. This is shown in Fig.~\ref{ref-fig14}a for the magnetic
subsystem and in the inset for the ferroelectric layers.

For the inner layers of the magnetic subsystem, as the parameter
$J_{mf}$ increases, the type of transition changes (Fig.~\ref{ref-fig14}b).

Phase diagram in Fig.~\ref{ref-fig16}a shows the effect of $J_{mf}$ on the transition
temperature of the interface magnetic and ferroelectric layers. One
can see that the transition temperature increases as the absolute
value of $J_{mf}$ increases. At $J_{mf}=-3$ and below the transition
temperatures for the magnetic and ferroelectric layers become distinct.

Phase diagram in Fig.~\ref{ref-fig16}b shows the effect of the external electric field $E$ on the transition
temperature of the interface magnetic and ferroelectric layers.
One finds that the transition temperature is almost unchanged when
we increase $E^z$ up to $E^z=0.5$. For large values of
$|J_{mf}|$ ($J_{mf}{\ll}-3$) the transition temperature is not sensitive to
$E^z$.

\begin{figure}
\vspace{-10pt}
\begin{center}
\includegraphics[scale=0.60]{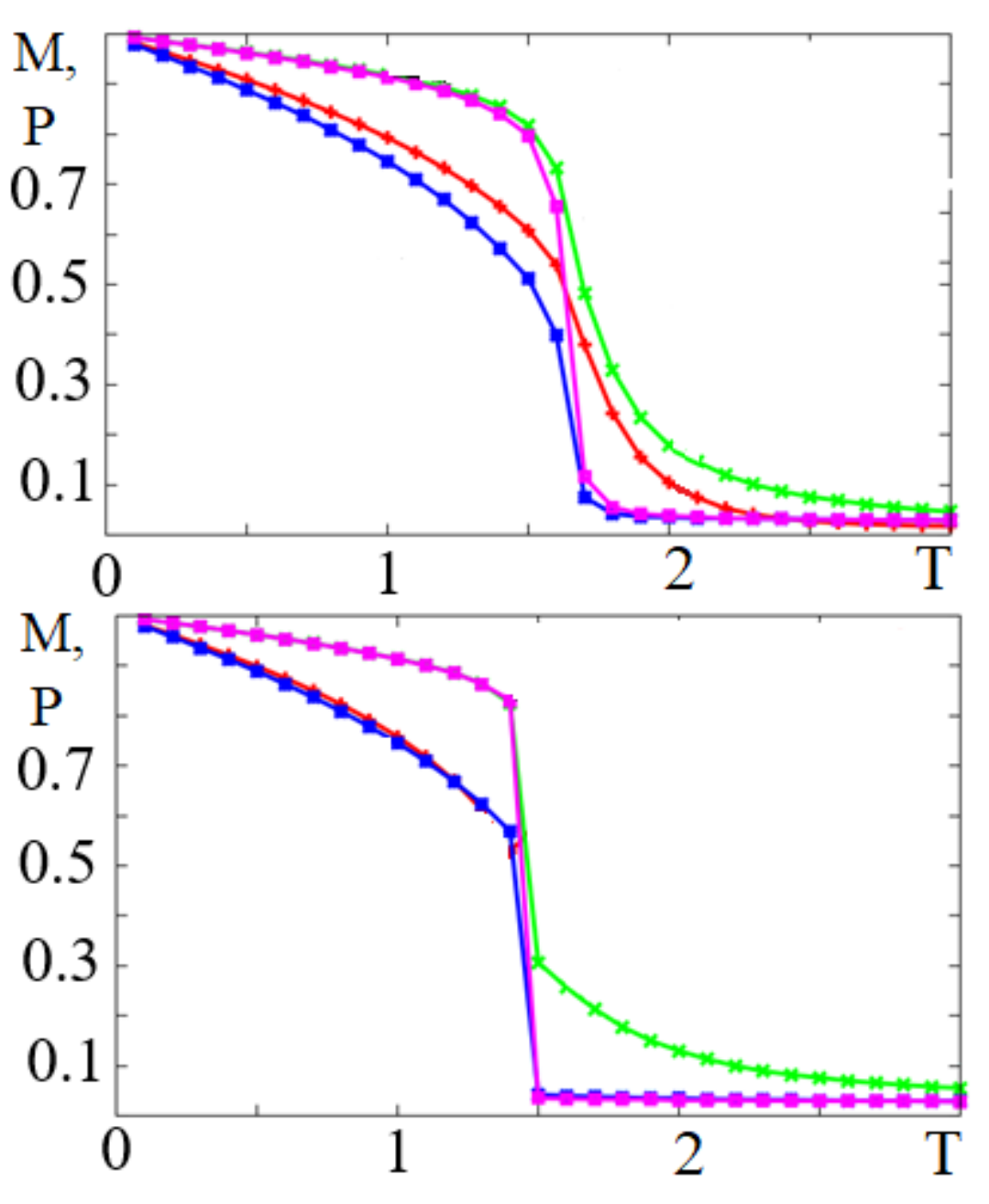}
\end{center}
\vspace{-10pt}
\caption {Temperature dependence of layer magnetizations and polarizations for $J^{m}=1$, $J^{f}=1$, $J_{mf}=-2.5$,
$H^z=0$, with $E^z=0$ (top) and $E^z=0.5$ (bottom). The coupling is given by  $H^{1}_{mf}$ (Eq. \ref{ham-sysme-3}). Color code: magenta (or purple) lines for the first layer and fourth magnetic layers (interface layers),
blue lines for the second and third (interior magnetic) layers.
Green lines for the  first and fourth (interface) ferroelectric layers,
red lines for the second and third interior ferroelectric layers,
respectively.
}\label{ref-fig18}
\end{figure}

\begin{figure}[h]
\vspace{10pt}
\begin{center}
\includegraphics[scale=0.45]{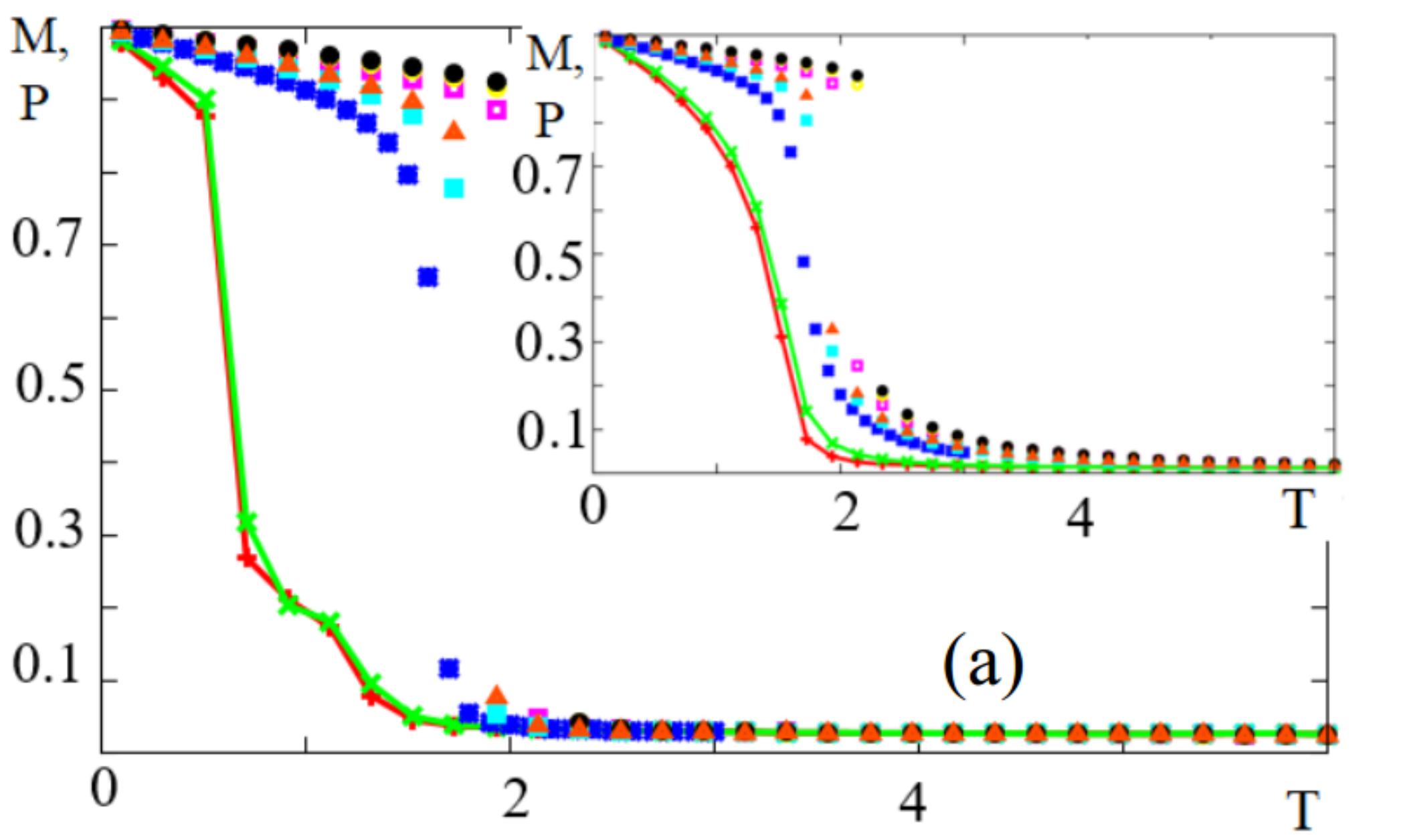}
\includegraphics[scale=0.43]{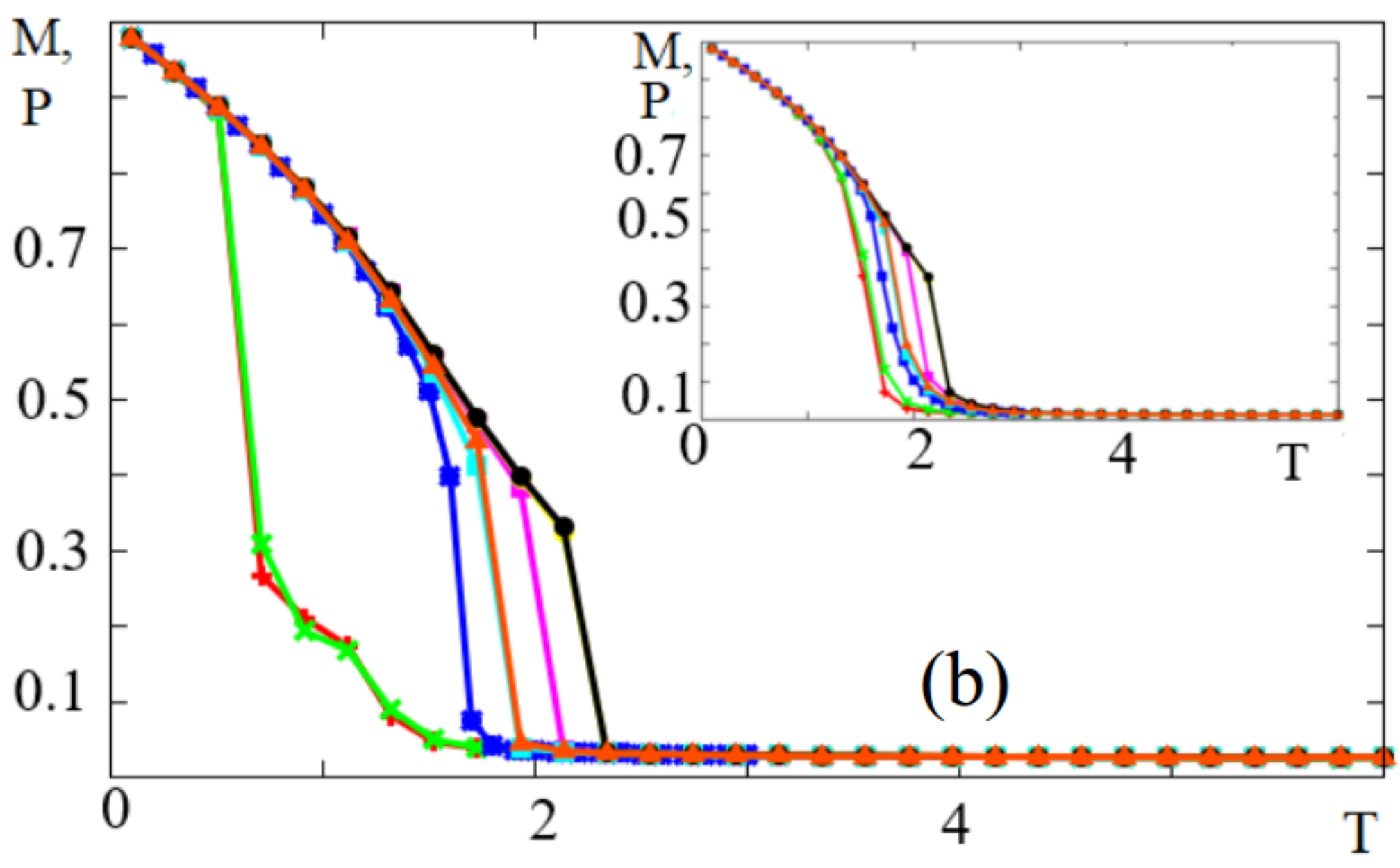}
\end{center}
\vspace{-20pt} \caption {(a) Temperature dependence of interface layers of
magnetic and ferroelectric films (inset) for $J^{m}=1$, $J^{f}=1$, $H^z=E^z=0$. Red line for the case $J_{mf}=-0.25$,
green line for $J_{mf}=-0.5$. Blue points: $J_{mf}=-2.5$, light blue points:
$J_{mf}=-3.5$, black points: $J_{mf}=-7$;  (b) Temperature dependence of interior layers of
magnetic and ferroelectric films (inset)  with the same parameters and color code. The coupling is given by $H^{1}_{mf}$.\label{ref-fig14}} \vspace{10pt}
\end{figure}

\begin{figure}
\begin{center}
\includegraphics[scale=0.55]{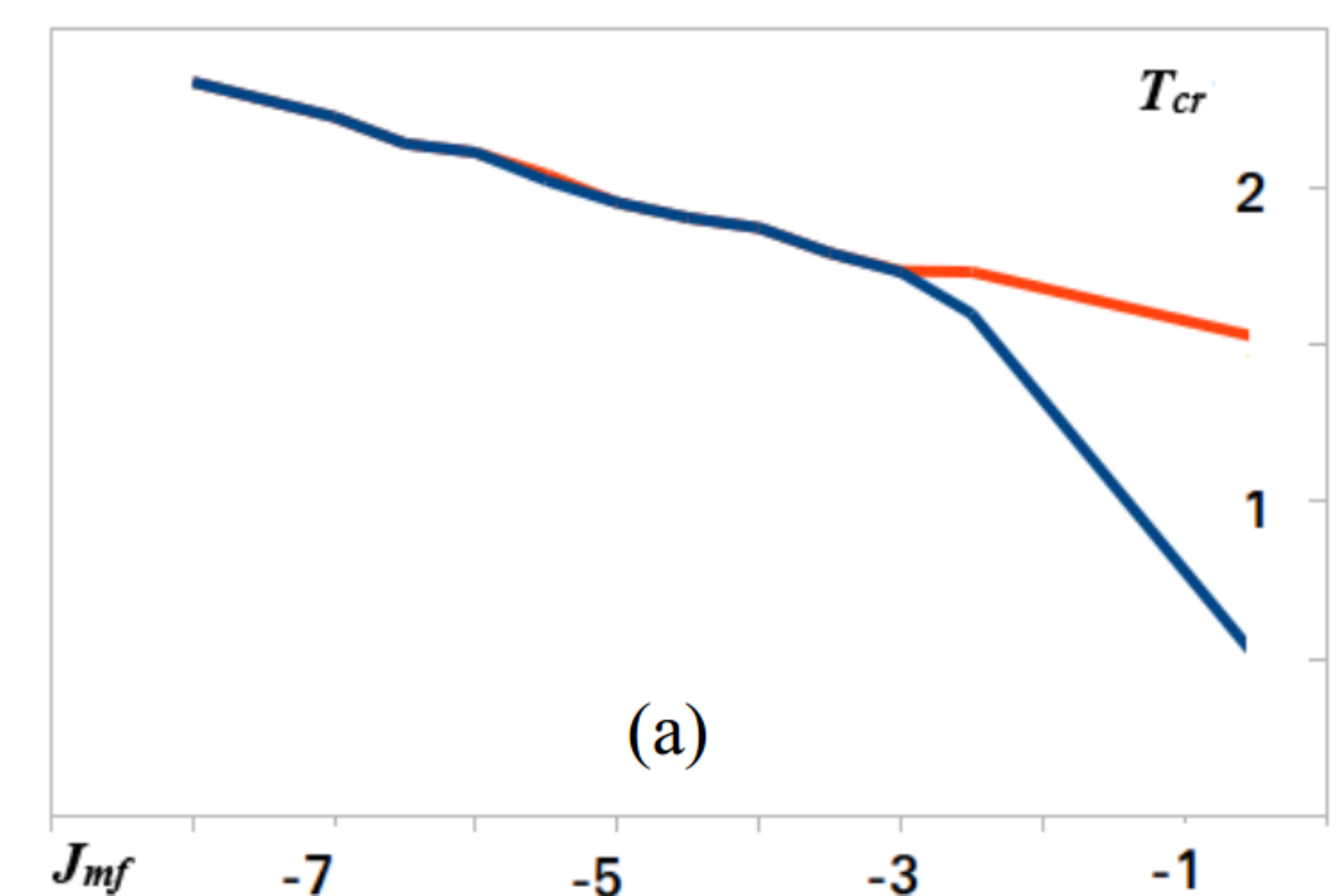}
\includegraphics[scale=0.47]{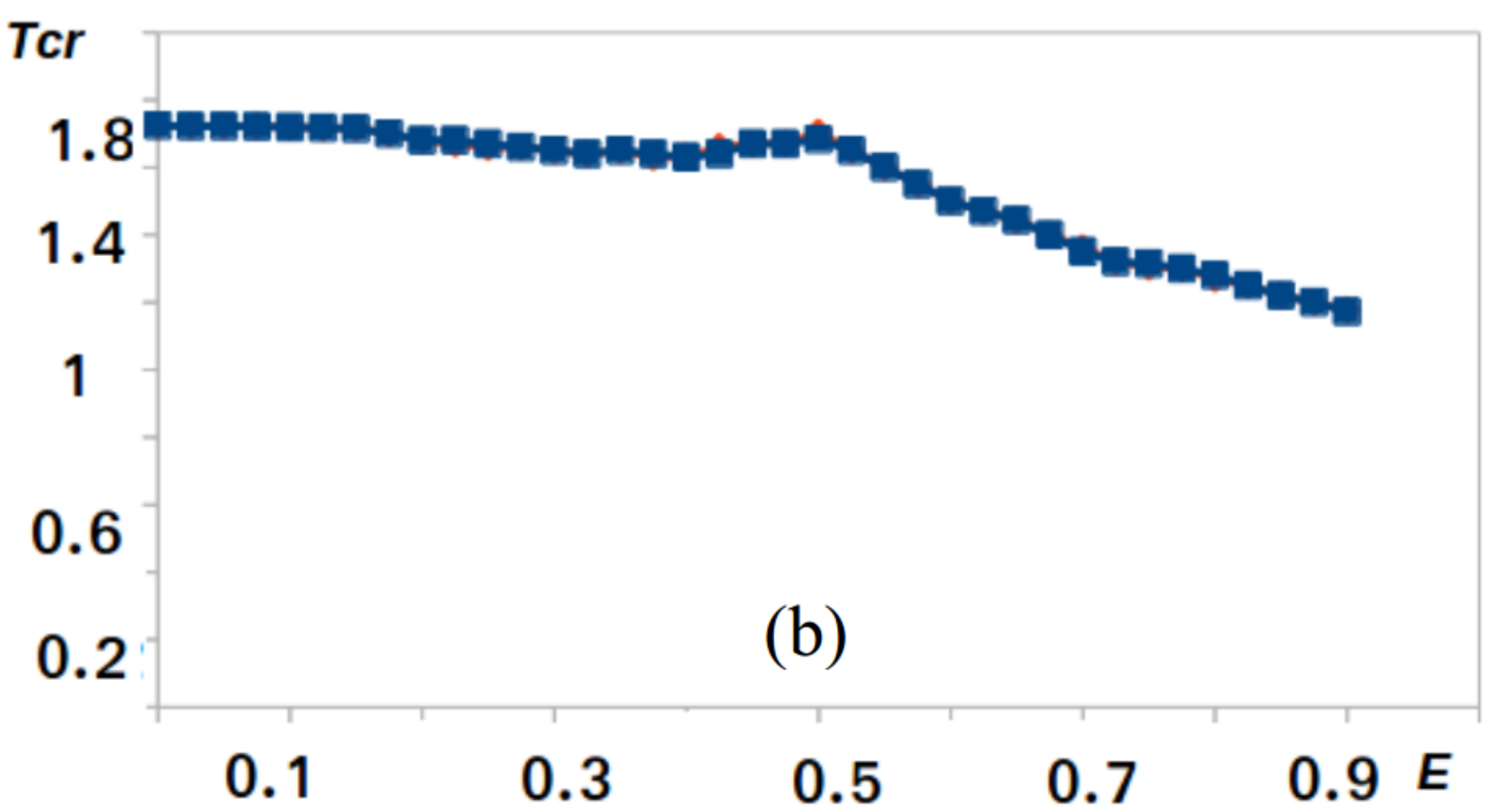}
\end{center}
\vspace{-10pt} \caption {(a) Phase diagram in $T-J_{mf}$ plane. Here
$J^{m}=1$, $J^{f} = 1$, $H^z=E^z=0$. Red line and black line
are for the ferroelectric critical temperature and
the magnetic critical temperature, respectively. The coupling is given by $H^{1}_{mf}$ ; (b) Phase diagram in $T-E$ plane ($E$ stands for $E^z$).
Here $J^{m}=1$, $J^{f} = 1$, $H^z=0$, $J_{mf}=-3.5$.}\label{ref-fig16} \vspace{-10pt}
\end{figure}

Figure ~\ref{ref-fig18} shows the effect of the competition between the
magnetoelectric interaction and the external electric field. With moderate
magnetoelectric interaction ($J_{mf}=-2.5$), we can
remark that the interface ferroelectric layer undergoes a second-order
phase transition at $T_{c}=1.77$, the magnetic layers undergo a
second-order phase transition at $T_{c} = 1.64$.
When we include an external electric field, both subsystems undergo
a first-order phase transition at the same temperature $T_{c} = 1.5$.

If the magnetoelectric interaction has a large value
and the external electric field is zero, we have seen above that the interface magnetic and ferroelectric
layers undergo a first-order phase transition. The inner magnetic layer undergoes a second-order
phase transition while the internal ferroelectric
layers are not subject to a phase transition. Now if we apply an
electric field for instance $E^z=0.5$, the inner ferroelectric layers
undergo a second-order phase transition (not shown).

%Fig11

%Fig12

%Fig13

To conclude this section, let us emphasize that beyond the two models for interface coupling studied above, the Dzyaloshinskii-Moriya interface interaction of the form $J_{mf}\vec P_k\cdot (\vec S_i\times  \vec S_j)$ may induce unexpected phenomena at the magneto-ferroelectric interface \cite{sergienko2006role,diep2018skyrmion}. Work is under way to investigate this coupling model.

\section{Mean-Field Theory}\label{MF}

Let us show some analytical results obtained by us using the mean-field (MF) theory for the Hamiltonian
\begin{equation}
\label{eq-ham-sys-1}
{\cal H}=H_{m}+H_{f}+H_{mf},
\end{equation}
where
\begin{equation}
\label{eq-ham-sys-2}
H_{mf}=-J_{mf}\sum_{i,j,k}{\vec{S}_{i}\cdot \vec{S}_{j}}P_{k},
\end{equation}
We consider the spin at the site $i$ and ferroelectric polarization at the site $l$.  We can write their local fields from the NN as

\begin{eqnarray}
H_{i}&=&-J^{m}\sum_{\vec{p}}\vec{S_{i}}\cdot \vec{S}_{i+p}
-\vec{H}\cdot \vec{S_{i}}-\frac{J_{mf}}{2}\delta_{i,4}P_{i+1}
\sum_{\vec{p}}\vec{S_{i}}\cdot \vec{S}_{i+p}\label{eq-mft1}\\
H_{l}&=&-J^{f}\sum_{\vec{p}}P_{l}P_{l+p}-E^zP_{l}
-\frac{J_{mf}}{2}\delta_{l,5}\sum_{\vec{p}}P_{l}\vec{S}_{l-1}\cdot \vec{S}_{l+1+p}\label{eq-mft2}
\end{eqnarray}
or
\begin{eqnarray}
H_{i}&=&-\bar{H}S^{z}_{i}\label{eq-mft3}\\
H_{l}&=&-\bar{H_{2}}P_{l}\label{eq-mft4}
\end{eqnarray}
here
\begin{eqnarray}
\bar{H}&=&\xi_{i}(\langle{S_z}\rangle+\langle{\Delta S_z}\rangle)+H^z\label{eq-mft5}\\
\bar{H_2}&=&J_{f}C_{3}(\langle{P_z}\rangle+\langle{\Delta P_z}\rangle)-\frac{J_{mf}}{2}\delta_{l,5}C_{2}S^{z}_{l-1}(\langle{S_z}\rangle+\langle{\Delta S_z}\rangle)+E^z \label{eq-mft6}\\
\xi_{i} &=&J_{m}C_{1}-\frac{J_{mf}}{2}\delta_{i,4}C_{2}P_{i+1}\label{eq-mft7}
\end{eqnarray}
where for notation convenience we write $P_z$ instead of $P$.

We choose the $z$ axis for the spin quantization axis. The average value of the $xy$ spin components are then zero since the spin precesses circularly around the $z$ axis:

\begin{equation}
\label{eq-mft9}
 \langle{S^{x}_{i+p}}\rangle=\langle{S^{y}_{i+p}}\rangle=\langle{S^{x}_{l-1+p}}\rangle=
 \langle{S^{y}_{l-1+p}}\rangle=0
\end{equation}

\begin{equation}
\label{eq-mft10}
\langle{S^{z}}\rangle+\langle{\Delta S^{z}}\rangle =\frac{\sum^{S}_{S^{z}_{i}=-S} S^{z}_{i}\exp(-\beta H_{i}) }{Z_{i}}
\end{equation}
where the partition function is
\begin{eqnarray}
Z_{i}&=&\sum^{S}_{S^{z}_{i}=-S}\exp(-\beta H_{i})=\sum^{S}_{S^{z}_{i}=-S} S^{z}_{i}\exp(-\beta \bar{H}S^{z}_{i})\label{eq-mft11}\\
Z_{i}&=&\frac{\sinh(\beta \bar{H}(S+\frac{1}{2}))}{\sinh(\frac{1}{2}\beta \bar{H})}\label{eq-mft12}
\end{eqnarray}
where
\begin{equation}
\label{eq-mft13}
S=|\vec{S_{i}}|
\end{equation}

We obtain
\begin{equation}
\label{eq-mft14}
\langle{S^{z}}\rangle+\langle{\Delta S^{z}}\rangle =B_{S}(\beta S \bar{H})
\end{equation}
here $B_{S}(\beta S \bar{H})$ is the Brillouin function defined by
\begin{equation}
\label{eq-mft15}
B_{S}(\beta S \bar{H})=\frac{2S+1}{2S}\coth(\frac{(2S+1)\beta S \bar{H}}{2S})-\frac{1}{2S}\coth(\frac{\beta S \bar{H}}{2S})
\end{equation}

If $H^{z}$ is very weak, we can suppose that $\langle{\Delta S^{z}}\rangle\rightarrow 0$ and in such a case we can expand the Brillouin function near $x_{0}=\beta\xi_{i}S\langle{S^{z}}\rangle$
\begin{equation}
\label{eq-mft16}
\langle{\Delta S^{z}}\rangle=\frac{S}{k_{B}T}(SH^{z}+S\xi_{i}\langle{\Delta S^{z}}\rangle)\frac{\partial B_{S}(x_0)}{\partial x_{0}}
\end{equation}

If $H^{z}=0$ then
\begin{equation}
\label{eq-mft17}
\langle{S^{z}}\rangle=SB_{S}(x_0)
\end{equation}

At high temperature $\beta\langle{S^{z}}\rangle\ll 1$ and
\begin{equation}
\label{eq-mft17}
B_S(\beta S \bar{H})\approx\frac{S+1}{3S}\beta S \bar{H}-\frac{[S^{2}+(S+1)^2](S+1)}{90S^3}(\beta S \bar{H})^3+
+O((\beta S \bar{H})^5)
\end{equation}
The previous equation becomes
\begin{equation}
\label{eq-mft18}
\langle{S^{z}}\rangle[\frac{\xi_{i}S(S+1)}{3k_{B}T}-1]=\frac{S(S+1)[S^{2}+(S+1)^{2}]}{90}(\frac{\xi_{i}}{k_{B}T})^3 (\langle{S^{z}}\rangle)^3
\end{equation}

This equation has a solution $\langle{S^{z}}\rangle\neq 0$ only if
\begin{equation}
\label{eq-mft19}
\frac{(2J_{m}C_{1}+J_{mf}\delta_{i,4}P_{i+1}C_{2})S(S+1)}{6k_{B}T}-1>0
\end{equation}
namely
\begin{equation}
\label{eq-mft20}
T<\frac{2J_{m}C_{1}+J_{mf}\delta_{i,4}P_{i+1}C_{2}S(S+1)}{6k_{B}}=T_c
\end{equation}
for $H_{l}$ we can write in the same manner the MF equations, and one can obtain for $\langle{P^{z}}\rangle+\langle{\Delta P^{z}}\rangle$
\begin{equation}
\label{eq-mft21}
\langle{P^{z}}\rangle+\langle{\Delta P^{z}}\rangle=PB_{P}(\beta P\bar{H_{2}})
\end{equation}
where $B_{P}(\beta P\bar{H_{2}})$ is the Brillouin function defined by
\begin{equation}
\label{eq-mft22}
B_{P}(\beta P\bar{H_{2}})=\frac{2P+1}{2P}coth(\frac{(2P+1)\beta P\bar{H_{2}}}{2P})-
\frac{1}{2P}coth(\frac{\beta P\bar{H_{2}}}{2P})
\end{equation}

In zero applied electric field we can write
\begin{equation}
\label{eq-mft23}
\langle{P^{z}}\rangle=PB_{P}(y_{0})
\end{equation}
here
\begin{eqnarray}
y_{0}&=&\frac{1}{k_{B}T}(J_{f}C_{1}P\langle{P^{z}}\rangle+y_{1}(\langle{S^{z}}\rangle)+\langle{\Delta S^{z}}\rangle)\label{eq-mft24}\\
y_{1}&=&J_{mf}\frac{P}{2}C_{2}\delta_{l,5}S^{z}_{l-1}\label{eq-mft25}
\end{eqnarray}
At high temperature, $\langle{P^{z}}\rangle=PB_{P}(y_{0})$
becomes
\begin{eqnarray}
\langle{P^{z}}\rangle&=&\frac{P+1}{k_{B}T}(J_{f}C_{3}P\langle{P^{z}}\rangle)+y_{1}(\langle{S^{z}}\rangle-\langle{\Delta S^{z}}\rangle)\nonumber\\
&&-\frac{(P^{2}+(P+1)^2)(P+1)}{90P^3}(\frac{J_{f}C_{3}P\langle{P^{z}}\rangle}{k_{B}T}+y_{1}(\langle{S^{z}}\rangle-\langle{\Delta S^{z}}\rangle)^3)\label{eq-mft26}
\end{eqnarray}

For our superlattice we can obtain for each layer the following system of equations
\begin{equation}
\label{eq-mft27}
\langle{S^{z}_{1,4}}\rangle+\langle{\Delta S^{z}_{1,4}}\rangle=SB_{\varphi_{1-4}}+SU_{\psi_{1-4}}
\end{equation}

\begin{equation}
\label{eq-mft28}
B_{\varphi_{1-4}}=B_{S}((\frac{J_{m}SC_{1}}{k_{B}T}+\frac{J_{mf}SC_{2}}{k_{B}T}\langle{P^{z}_{5,8}}\rangle)\langle{S^{z}_{1,4}}\rangle+\frac{J_{m}S}{k_{B}T}\langle{S^{z}_{2,3}}\rangle)
\end{equation}

\begin{equation}
\label{eq-mft29}
U_{\psi_{1-4}}=(\frac{J_{m}SC_{1}}{k_{B}T}+\frac{J_{mf}SC_{2}}{k_{B}T}\langle{P^{z}_{5,8}}\rangle)\langle{\Delta S^{z}_{1,4}}\rangle+\frac{HS^{2}}{k_{B}T})\frac{\partial B_{S}(u)}{\partial u}
\end{equation}

\begin{equation}
\label{eq-mft30}
u=(\frac{J_{m}SC_{1}}{k_{B}T}+\frac{J_{mf}SC_{2}}{k_{B}T}\langle{P^{z}_{5,8}}\rangle)\langle{S^{z}_{1,4}}\rangle
\end{equation}

\begin{equation}
\label{eq-mft31}
\langle{S^{z}_{2}}\rangle+\langle{\Delta S^{z}_{2}}\rangle=SB_{\varphi_{2}}+SU_{\psi_{2}}
\end{equation}
\begin{equation}
\label{eq-mft33}
B_{\varphi_{2}}=B_{S}(\frac{J_{m}SC_{1}\langle{S^{z}_{2}}\rangle}{k_{B}T}+\frac{J_{mf}S\langle{S^{z}_{1,4}}\rangle\langle{P^{z}_{7,8}}\rangle}{k_{B}T}+\frac{J_{m}S\langle{S^{z}_{3}}\rangle}{k_{B}T})
\end{equation}
\begin{equation}
\label{eq-mft34}
U_{\psi_{2}}=-S^2(\frac{J_{m}C_{1}\langle{\Delta S^{z}_{2}}\rangle}{k_{B}T}+\frac{H}{k_{B}T})\frac{\partial B_{S}(\frac{J_{m}SC_{1}\langle{S^{z}_{1,4}}\rangle}{k_{B}T})}{\partial (\frac{J_{m}SC_{1}\langle{S^{z}_{1,4}}\rangle}{k_{B}T})}
\end{equation}
\begin{equation}
\label{eq-mft35}
\langle{S^{z}_{3}}\rangle+\langle{\Delta S^{z}_{3}}\rangle=SB_{\varphi_{3}}+SU_{\psi_{3}}
\end{equation}
\begin{equation}
\label{eq-mft36}
B_{\varphi_{3}}=B_{S}(\frac{J_{m}SC_{1}\langle{S^{z}_{3}}\rangle}{k_{B}T}+\frac{J_{mf}S\langle{S^{z}_{2}}\rangle\langle{P^{z}_{7,8}}\rangle}{k_{B}T}+\frac{J_{m}S\langle{S^{z}_{1-4}}\rangle}{k_{B}T})
\end{equation}
\begin{equation}
\label{eq-mft37}
U_{\psi_{3}}=-S^2(\frac{J_{m}C_{1}\langle{\Delta S^{z}_{3}}\rangle}{k_{B}T}+\frac{H}{k_{B}T})\frac{\partial B_{S}(\frac{J_{m}SC_{1}\langle{S^{z}_{1,4}}\rangle}{k_{B}T})}{\partial (\frac{J_{m}SC_{1}\langle{S^{z}_{1,4}}\rangle}{k_{B}T})}
\end{equation}
\begin{equation}
\label{eq-mft38}
\langle{P^{z}_{5,8}}\rangle+\langle{\Delta P^{z}_{5,8}}\rangle=PB_{\varphi_{5,8}}+PU_{\psi_{5,8}}
\end{equation}
\begin{equation}
\label{eq-mft38}
B_{\varphi_{5,8}}=B_{P}(\frac{J_{f}PC_{3}\langle{P^{z}_{5,8}}\rangle}{k_{B}T}+\frac{J_{mf}PC_{2}(\langle{S^{z}_{1,4}}\rangle)^2}{k_{B}T}+\frac{J_{f}P\langle{P^{z}_{6}}\rangle}{k_{B}T})
\end{equation}
\begin{equation}
\label{eq-mft39}
U_{\psi_{5,8}}=-P^{2}(\frac{J_{f}C_{3}\langle{\Delta P^{z}_{5,8}}\rangle}{k_{B}T}+\frac{J_{mf}C_{2}(\langle{S^{z}_{1,4}}\rangle)^2}{k_{B}T}+\frac{E}{k_{B}T})\frac{\partial B_{P}(v)}{\partial v}
\end{equation}
\begin{equation}
\label{eq-mft40}
v=\frac{J_{f}PC_{3}\langle{P^{z}_{5,8}}\rangle}{k_{B}T}+\frac{J_{mf}PC_{2}(\langle{S^{z}_{1,4}}\rangle)^2}{k_{B}T}
\end{equation}
\begin{equation}
\label{eq-mft41}
\langle{P^{z}_{6}}\rangle+\langle{\Delta P^{z}_{6}}\rangle=PB_{\varphi_{6}}+PU_{\psi_{6}}
\end{equation}

\begin{equation}
\label{eq-mft42}
B_{\varphi_{6}}=B_{P}(P(\frac{J_{f}C_{3}\langle{P^{z}_{6}}\rangle}{k_{B}T}+\frac{J_{mf}P(\langle{S^{z}_{1,4}}\rangle)^2}{k_{B}T}+\frac{J_{f}\langle{P^{z}_{7}}\rangle}{k_{B}T}))
\end{equation}

\begin{equation}
\label{eq-mft43}
U_{\psi{6}}=-P^2(\frac{J_{f}C_{3}\langle{\Delta P^{z}_{6}}\rangle}{k_{B}T}+\frac{E}{k_{B}T})\frac{\partial B_{P}(\frac{J_{f}PC_{3}}{k_{B}T})}{\partial (\frac{J_{f}PC_{3}}{k_{B}T})}
\end{equation}

\begin{equation}
\label{eq-mft44}
\langle{P^{z}_{7}}\rangle+\langle{\Delta P^{z}_{7}}\rangle=PB_{\varphi{7}}+PU_{\psi_{7}}
\end{equation}

\begin{equation}
\label{eq-mft45}
B_{\varphi{7}}=B_{P}(\frac{J_{f}PC_{3}\langle{P^{z}_{7}}\rangle}{k_{B}T}+\frac{J_{mf}P(\langle{S^{z}_{1,4}}\rangle)^2}{k_{B}T}+\frac{J_{f}P\langle{P^{z}_{6}}\rangle}{k_{B}T})
\end{equation}

\begin{equation}
\label{eq-mft46}
U_{\psi{7}}=-P^2(\frac{J_{f}C_{3}\langle{\Delta P^{z}_{7}}\rangle}{k_{B}T}+\frac{E}{k_{B}T})\frac{\partial B_{P}(\frac{J_{f}PC_{3}}{k_{B}T})}{\partial (\frac{J_{f}PC_{3}}{k_{B}T})}
\end{equation}

\begin{figure}[h]
\vspace{10pt}
\begin{center}
\includegraphics[scale=0.46]{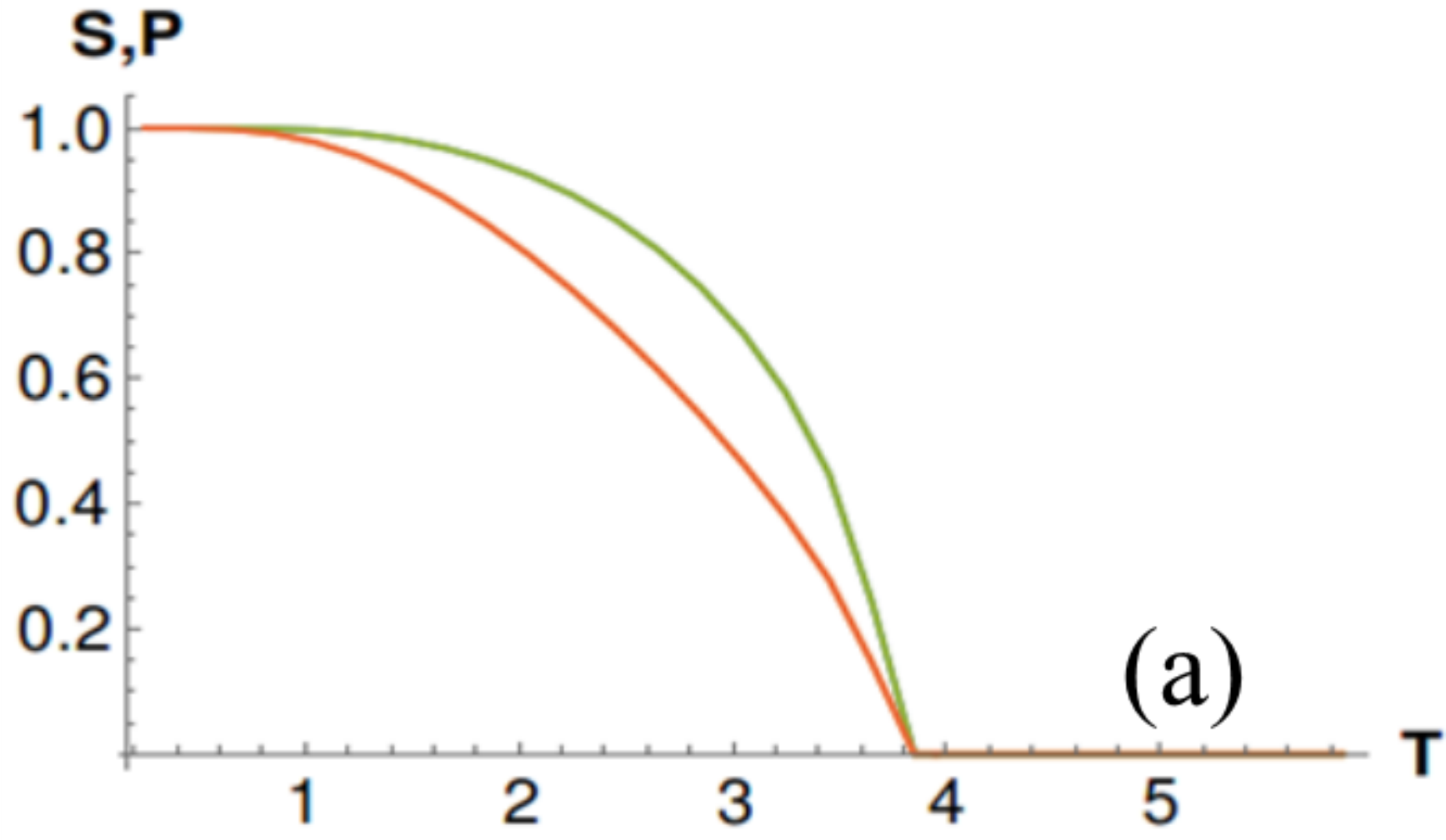}
\includegraphics[scale=0.46]{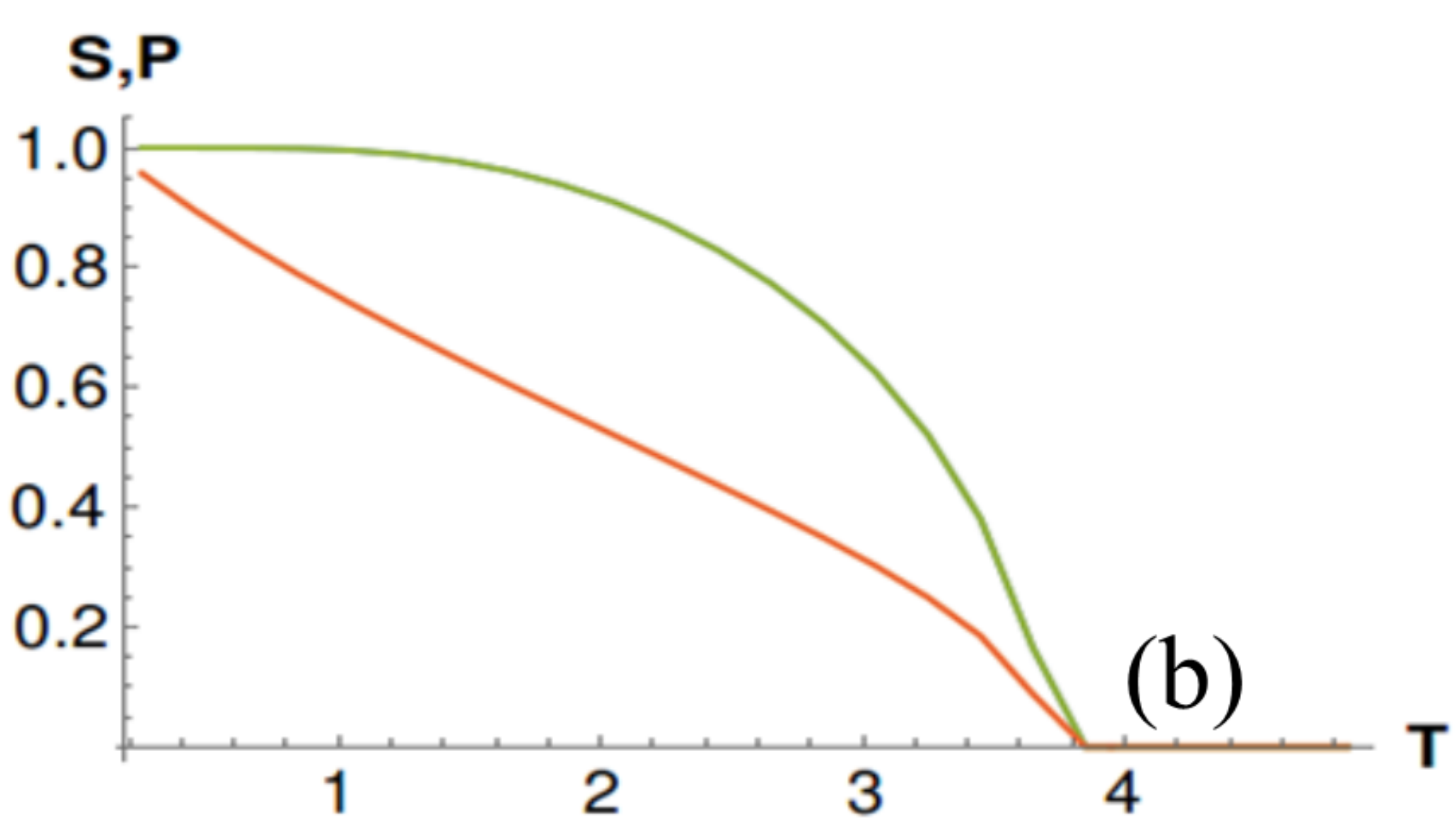}
\includegraphics[scale=0.46]{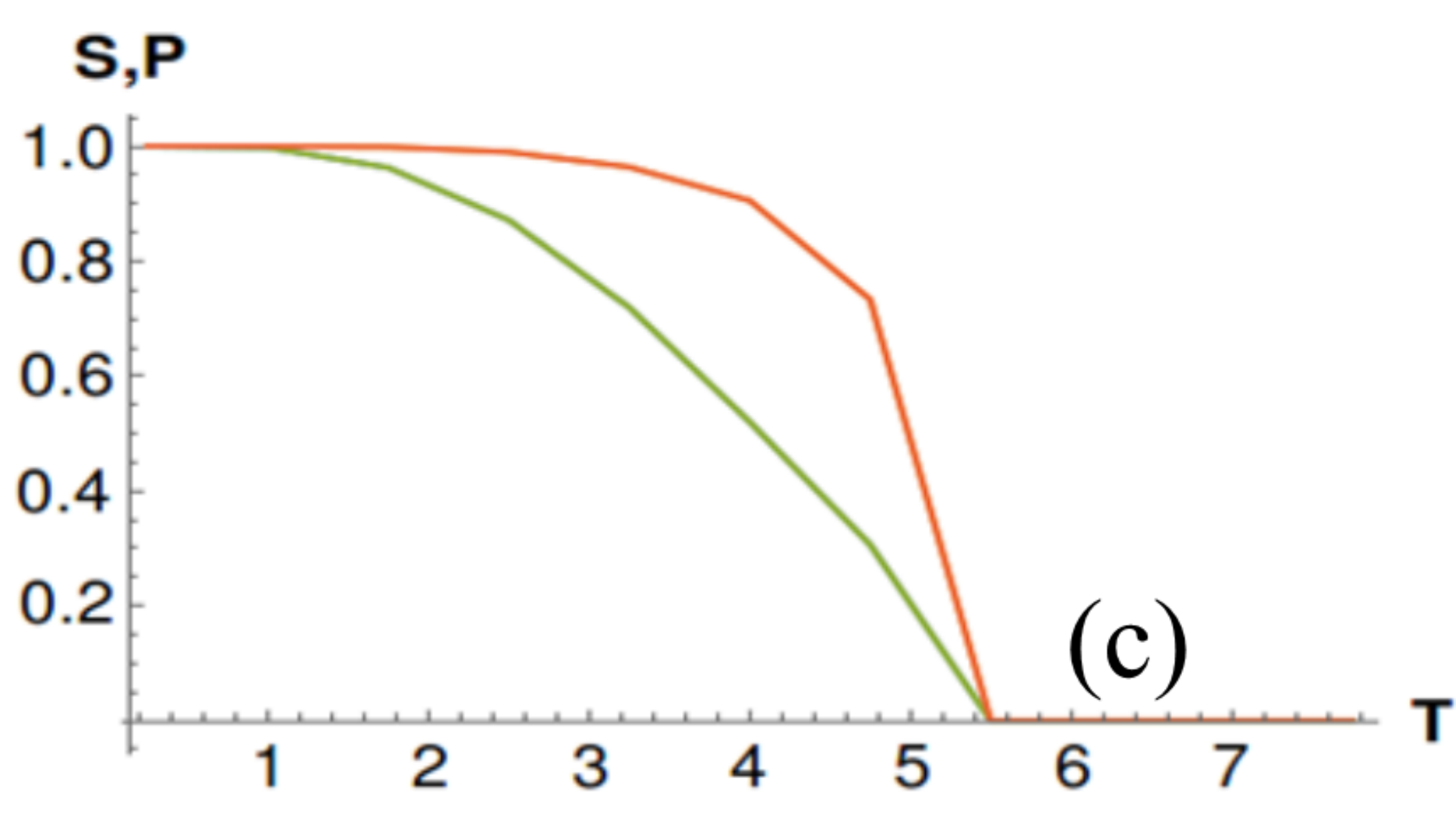}
\end{center}
\vspace{10pt} \caption{Mean-field results with
$J_{mf1}=J_{mf2}\equiv J_{mf}$  (a) Temperature dependence of the magnetization and
polarization in the case $J^{m}=1$, $J^{f}=1$,
$J_{mf}=-0.12$, $H^z=E^z=0$. Red lines for the interface
$P$ and $M$, green line for the inner layers $M_{2,3}$ and $P_{2,3}$; (b) Temperature dependence of the magnetization
and polarization in the case $J^{m}=1$, $J^{f}=1$,
$J_{mf}=-0.55$, $H^{z}=0$ and $E^{z}=0$. Red lines for the interface
$P$ and $M$, green lines for the inner layers $M_{2,3}$ and $P_{2,3}$; (c)  Temperature dependence of the magnetization and polarization in the case
$J^{m}=1$, $J^{f}=1$, $J_{mf}=-0.85$, $H^{z}=E^{z}=0$. Red
lines for the interface $P$ and $M$, green lines for the inner layers
$M_{2,3}$ and $P_{2,3}$.  $|\vec{S}|=|P|=1$.} \label{ref-fig20} \vspace{-10pt}
\end{figure}

Figure ~\ref{ref-fig20} shows the effect of the magnetoelectric interaction on
 the temperature dependence of the polarization and the magnetization, for both the interface and the inner layer.
In the MF theory, the magnetization and ferroelectric polarization coincide if their amplitudes are the same.  This is because the $xy$ spin components are neglected, making Heisenberg spins $S$ equivalent to Ising spins $P$.

%Fig14

If we compare Figs.  ~\ref{ref-fig20} and  ~\ref{ref-fig2}-\ref{ref-fig4} one can see an agreement between MC and MF theory that the interface order parameter depends strongly on the interface coupling and have different value from that of the interior layers.

\begin{figure}[h]
\vspace{-10pt}
\begin{center}
\includegraphics[scale=0.50]{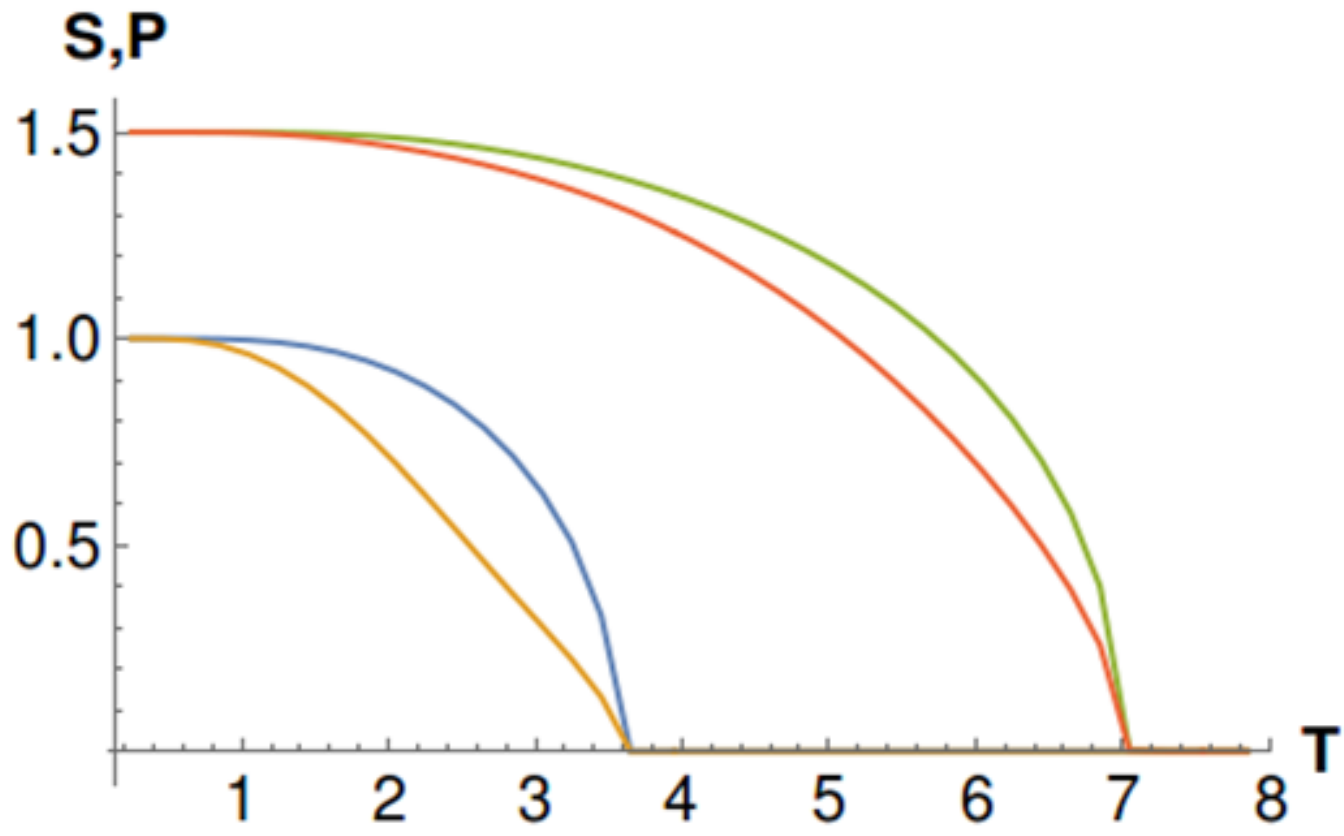}
\end{center}
\vspace{10pt} \caption{Temperature dependence of the magnetization and
polarization for $|\vec{S}|=1$, $|P|=1.5$.  $J^{m}=-1$, $J^{f}=1$,
$J_{mf}=-0.12$, $H^z=E^z=0$. Red lines for the interface
$P$,  gold line for interface $M$, blue line for inner layers $M_{2,3}$ and green line
for inner layers $P_{2,3}$.} \label{ref-fig27} \vspace{10pt}
\end{figure}

If we take $|P|=1.5$ we see different transition temperatures
for magnetic and ferroelectric films as seen in Figure ~\ref{ref-fig27}.

%Fig15

\section{Conclusion}\label{sect-conc}

We have studied in this paper the effects of the temperature,
external magnetic and electric fields, the magnetoelectric coupling
in a multiferroic superlattice
formed by alternating magnetic and ferroelectric films. Magnetic
films in this work were modeled as films of simple cubic lattice with
Heisenberg spins. Electrical polarizations of values $\pm{1}$ were
assigned at each lattice site in the ferroelectric
films.

We have studied these superlattices with MC simulations and with a MF theory. Various physical quantities have been obtained to identify and characterize the phase transition in each subsystem as functions of temperature $T$, interface coupling parameter and applied magnetic and electric fields. Two models of interface coupling have been considered.  The MC and MF calculations agree with each other with regard to the interface order parameters.

Among our MC results let us mention the change of the nature of the phase transition when the interface coupling parameter changes.  Various phase diagrams have been established which show that magnetic and ferroelectric phase transitions are closely connected. The interface magnetic and ferroelectric layers have distinct behaviors compared to the inner layers. This is known when there is a loss of translation invariance such as the presence of an impurity, a surface or an interface.

We have worked out a laborious mean-field formalism for superlattices. The application of this in this paper was intentionally limited, but there are wider applications in many system geometries and in various interacting films such as ferri-electric superlattices and frustrated superlattices which have not been considered here.

To conclude, let us emphasize that we have studied in  this work two models for interface coupling. Other models of interface magneto-ferroelectric coupling such as the Dzyaloshinskii-Moriya interaction may induce unexpected phenomena at the magneto-ferroelectric interface. Work is under way to investigate this coupling model.

\section*{Acknowledgment}
One of us (IFS) wishes to thank Campus France for a financial support (contract P678172A) during the course of the present work.

\section*{References}
\label{sect-ref}
\bibliographystyle{elsarticle-num}
\bibliography{bibliography}

\begin{thebibliography}{10}
\expandafter\ifx\csname url\endcsname\relax
  \def\url#1{\texttt{#1}}\fi
\expandafter\ifx\csname urlprefix\endcsname\relax\def\urlprefix{URL }\fi
\expandafter\ifx\csname href\endcsname\relax
  \def\href#1#2{#2} \def\path#1{#1}\fi

\bibitem{diep:hal-01084599}
H.~T. Diep, {Theory Of Magnetissm - Application to Surface Physics}, {World
  Scientific}, 2014.

\bibitem{Diep201631}
H.~T. Diep, Theoretical methods for understanding advanced magnetic materials:
  The case of frustrated thin films, Journal of Science: Advanced Materials and
  Devices 1~(1) (2016) 31 -- 44.

\bibitem{Prudnikov2015}
V.~V. Prudnikov, P.~V. Prudnikov, D.~E. Romanovskii, Monte carlo simulation of
  multilayer magnetic structures and calculation of the magnetoresistance
  coefficient, JETP Letters 102~(10) (2015) 668--673.

\bibitem{Ramazanov2016}
M.~K. Ramazanov, A.~K. Murtazaev, Phase transitions in the antiferromagnetic
  layered ising model on a cubic lattice, JETP Letters 103~(7) (2016) 460--464.

\bibitem{kamilov1999}
I.~K. Kamilov, A.~K. Murtazaev, K.~K. Aliev, Monte carlo studies of phase
  transitions and critical phenomena, Physics-Uspekhi 42~(7) (1999) 689--709.

\bibitem{pyatakov}
A.~P. Pyatakov, A.~K. Zvezdin, Magnetoelectric and multiferroic media,
  Physics-Uspekhi 55~(6) (2012) 557--581.

\bibitem{PhysRevLett}
Y.~Weng, L.~Lin, E.~Dagotto, S.~Dong, Inversion of ferrimagnetic magnetization
  by ferroelectric switching via a novel magnetoelectric coupling, Phys. Rev.
  Lett. 117 (2016) 037601.

\bibitem{Iijima_1992}
K.~Iijima, T.~Terashima, Y.~Bando, K.~Kamigaki, H.~Terauchi, Atomic layer
  growth of oxide thin films with perovskite‐type structure by reactive
  evaporation, Journal of Applied Physics 72~(7).

\bibitem{Oneill2000}
D.~O’Neill, R.~M. Bowman, J.~M. Gregg, Dielectric enhancement and
  maxwell–wagner effects in ferroelectric superlattice structures, Applied
  Physics Letters 77~(10).

\bibitem{PhysRevB.55.11218}
B.~D. Qu, W.~L. Zhong, R.~H. Prince, Interfacial coupling in ferroelectric
  superlattices, Phys. Rev. B 55 (1997) 11218--11224.

\bibitem{ramesh}
R.~Ramesh, N.~A. Spaldin, Multiferroics: progress and prospects in thin films,
  Nature materials 6~(1) (2007) 21--29.

\bibitem{PhysRevB.85.184413}
Y.~Magnin, H.~T. Diep, Monte carlo study of magnetic resistivity in
  semiconducting mnte, Phys. Rev. B 85 (2012) 184413.

\bibitem{Kharrasov2016}
M.~K. Kharrasov, I.~R. Kyzyrgulov, I.~F. Sharafullin, A.~G. Nugumanov, Phase
  transitions and critical phenomena in multiferroic films with orthorhombic
  magnetic structure, Bulletin of the Russian Academy of Sciences: Physics
  80~(6) (2016) 695--697.

\bibitem{PhysRevB.73.094434}
I.~A. Sergienko, E.~Dagotto, Role of the dzyaloshinskii-moriya interaction in
  multiferroic perovskites, Phys. Rev. B 73 (2006) 094434.

\bibitem{lamekhov2015}
I.~V. Bychkov, D.~A. Kuzmin, V.~G. Shavrov, S.~Lamekhov, Monte carlo modelling
  of two dimensional multiferroics, in: Achievements in Magnetism, Vol. 233 of
  Solid State Phenomena, Trans Tech Publications, 2015, pp. 379--382.

\bibitem{PhysRevB.87.094416}
P.~M. Leufke, R.~Kruk, R.~A. Brand, H.~Hahn, \textit{In situ} magnetometry
  studies of magnetoelectric lsmo/pzt heterostructures, Phys. Rev. B 87 (2013)
  094416.

\bibitem{Ort2014}
H.~H. Ortiz-Alvarez, C.~M. Bedoya-Hincapie, E.~Restrepo-Parra, Monte carlo
  simulation of charge mediated magnetoelectricity in multiferroic bilayers,
  Physica B: Condensed Matter 454 (2014) 235 -- 239.

\bibitem{0953-8984-14-27-310}
V.~Y. Irkhin, A new mechanism of first-order magnetization in multisublattice
  rare-earth compounds.

\bibitem{PhysRevB.66.014437}
L.~Gontchar, A.~Nikiforov, Superexchange interaction in insulating manganites r
  1- x a x mno 3 (x= 0, 0. 5), Physical Review B 66~(1) (2002) 014437.

\bibitem{PhysRevB.55.12408}
Z.~Zhang, et~al., Spin waves in several heisenberg systems: Three-sublattice
  with different exchange constants (j (ab)= j (bc) not equal j (ca)) and a
  superlattice with the elementary unit of four or three different layers.

\bibitem{Diepbook2}
H.~T. Diep, H.~Giacomini, Frustration - exactly solved frustrated models, in:
  Frustrated Spin Systems, World Scientific, 2013, pp. 1--58.

\bibitem{WANG2017104}
W.~Wang, F.-l. Xue, M.-z. Wang, Compensation behavior and magnetic properties
  of a ferrimagnetic mixed-spin (1/2, 1) ising double layer superlattice,
  Physica B: Condensed Matter 515 (2017) 104--111.

\bibitem{Fer2016}
A.~Feraoun, A.~Zaim, M.~Kerouad, Quantum monte carlo study of the electric
  properties of a ferroelectric superlattice, Solid State Communications 248
  (2016) 88 -- 96.

\bibitem{landau2014guide}
D.~P. Landau, K.~Binder, A guide to Monte Carlo simulations in statistical
  physics, Cambridge university press, 2014.

\bibitem{phu2009crossover}
X.~T.~P. Phu, V.~T. Ngo, H.~T. Diep, Crossover from first-to second-order
  transition in frustrated ising antiferromagnetic films, Physical Review E
  79~(6) (2009) 061106.

\bibitem{phu2009critical}
X.~T.~P. Phu, V.~T. Ngo, H.~T. Diep, Critical behavior of magnetic thin films,
  Surface Science 603~(1) (2009) 109--116.

\bibitem{PhysRevB.91.014436}
H.~T. Diep, Quantum theory of helimagnetic thin films, Phys. Rev. B 91 (2015)
  014436.

\bibitem{sergienko2006role}
I.~A. Sergienko, E.~Dagotto, Role of the dzyaloshinskii-moriya interaction in
  multiferroic perovskites, Physical Review B 73~(9) (2006) 094434.

\bibitem{diep2018skyrmion}
H.~T. Diep, S.~El~Hog, A.~Bailly-Reyre, Skyrmion crystals: Dynamics and phase
  transition, AIP Advances 8~(5) (2018) 055707.

\end{thebibliography}
\end{document}